\documentclass{aa}  

\usepackage{graphicx}
\usepackage{txfonts}
\begin{document}

\title{Two stellar clocks, one history: a Gaia colour–magnitude-diagram reconstruction of the solar-neighbourhood star-formation history tested against white dwarfs}

% \subtitle{XXXX}

\author{
M. Cignoni\inst{1,2,3}
\and
P. G. Prada Moroni\inst{1,2}
\and
M. Gabbanini\inst{1} 
\and
S. Degl'Innocenti\inst{1,2}
\and
A. Tinarelli\inst{1}
}

\institute{
Dipartimento di Fisica, Università di Pisa, Largo Bruno Pontecorvo 3, 56127 Pisa, Italy
\and
INFN, Largo Bruno Pontecorvo 3, 56127 Pisa, Italy
\and
INAF - Osservatorio di Astrofisica e Scienza dello Spazio di Bologna, Via Piero Gobetti 93/3, 40129 Bologna, Italy
}

\authorrunning{Cignoni et al.}
\titlerunning{The solar-neighbourhood SFH from Gaia and the WDLF}

%   \date{Received September 15, 1996; accepted March 16, 1997}

\abstract
  % context heading (optional)
   {The solar-neighbourhood star-formation history (SFH) can be constrained
   from the colour--magnitude diagram (CMD) of non-degenerate stars and tested
   independently with the white-dwarf luminosity function (WDLF). These two
   tracers sample complementary evolutionary phases and provide a consistency
   check on both the recovered SFH and the adopted WD cooling models.}
  % aims heading (mandatory)
   {We reconstruct the local SFH from the \emph{Gaia} DR3 CMD and test whether
   the resulting history reproduces the observed 40~pc WDLF.}
  % methods heading (mandatory)
   {We fit non-degenerate stars in a cylinder of radius 200~pc and vertical
   extent $\pm40$~pc using the \textsc{SFERA} synthetic-population code. The fit
   is performed with the PISA and PARSEC stellar libraries. Each CMD-derived
   solution is then used to generate a synthetic WD population for comparison
   with the 40~pc WDLF, adopting the same initial--final mass relation and four
   WD cooling model sets: BaSTI, MIST, La Plata, and Montr\'eal.}
  % results heading (mandatory)
  {The two CMD-based solutions agree on the main features of the local SFH:
low activity at the oldest ages, an early episode around 9--11~Gyr ago,
reduced activity between about 4 and 9~Gyr, and enhanced star formation over
the last few Gyr. The predicted WDLFs reproduce the bright and intermediate
part of the observed distribution for all cooling grids, but differ
substantially around the WDLF maximum and along the faint tail. MIST and
BaSTI shift too much weight to magnitudes fainter than the observed maximum,
whereas La Plata gives the closest match to the faint tail but leaves larger
residuals around the maximum. Montr\'eal provides the best overall compromise.
On this cooling scale, progressively removing the oldest synthetic WDs shows
that the faint tail disfavours a substantial contribution from WDs with total
ages older than about 11~Gyr.}

  % conclusions heading (optional), leave it empty if necessary 
%   {}

        \maketitle
%
%-------------------------------------------------------------------

%%%%%%%%%%%%%%%%% BODY OF PAPER %%%%%%%%%%%%%%%%%%
\section{Introduction}
Galaxy formation proceeds through both \emph{in situ} and \emph{ex situ} channels.
The former reflects internally regulated star formation, shaped by the
availability of cold gas, the local environment, and feedback from supernovae
and stellar winds. The latter encompasses accretion and mergers with other
systems, the capture of satellites, and the broader cosmic environment.
Disentangling the relative roles of these two processes requires knowledge of
\emph{stellar ages} across a galaxy's lifetime. Because ages cannot be measured
directly, they must be inferred from stellar models, and their accuracy depends
critically on precise distances \citep[see e.g.][]{soderblom2010,
soderblom2015}.

For decades, the lack of reliable parallaxes for large samples of Milky Way
stars severely limited attempts to reconstruct the Galaxy's evolutionary
history. This situation improved significantly with the Hipparcos mission
(\citealt{Hipparcos1997}) and was transformed by the advent of the \emph{Gaia}
mission (Data Release 3; DR3; June 2022, \citealt{GaiaDR3}), which has provided
accurate parallaxes and homogeneous photometry for vast numbers of stars across
the Galaxy. In doing so, \emph{Gaia} has effectively made it possible to
study the Milky Way as a galaxy in its own right, with the same tools
routinely applied to external systems.

In this context, the colour-magnitude diagram (CMD) of the non-degenerate
stellar population provides a powerful tool to infer the local star-formation
history (SFH). This is especially true for \emph{Gaia}, whose parallaxes and
photometry allow the oldest main-sequence turn-offs (oMSTOs) to be reached in
statistically significant stellar samples, thereby enabling direct constraints
on the full age range of Galactic stellar populations.

Strictly speaking, however, what is derived from the CMD is not the pristine star-formation rate of the solar neighbourhood at birth, but the \textit{dynamically evolved SFH}: the imprint left on the present-day population after billions of years of vertical heating and radial migration (see e.g. \citealt{Frankel2018}). This subtle but important distinction has been recognised since the earliest CMD-based studies of the solar neighbourhood.

The first CMD-based reconstructions relied on Hipparcos parallaxes, which provided precise distances for nearby stellar samples in the solar vicinity. Using Hipparcos data, \citet{Bertelli2001} and \citet{Schroeder2003} derived local SFHs that gradually increase toward recent times. \citet{Hernandez2000} focused instead on the last 3 Gyr, finding an overall declining trend modulated by finer temporal structure, with quasi-periodic variations on sub-Gyr timescales. \citet{vergely2002} and \citet{cignoni2006} likewise found SFHs characterised by a gradual rise in the star-formation rate, with the main peak occurring about 1.6 Gyr ago and 2-3 Gyr ago, respectively. Although these pioneering works set the stage for SFH studies in the solar vicinity, the Hipparcos CMDs remained sparsely populated, so the entire reconstruction was affected by small-number statistics; the oldest epochs were especially uncertain because the oMSTO was reached only in a very limited volume and was therefore sampled by too few stars to yield robust constraints.

Several groups have since revisited the problem using \emph{Gaia}, yet their results still differ in key respects. Using \emph{Gaia} DR1/TGAS data combined with Tycho-2 and APASS photometry, \citet{bernard2018} derived a nearly constant SFH over most of the past 10 Gyr, with only a modest enhancement in the last $\sim 4$ Gyr. Using local samples from \emph{Gaia} DR2, \citet{daltio2021} identified a period of enhanced activity between 2 and 5 Gyr ago, with lower star-formation activity at both older and younger ages, while \citet{alzate2021} inferred an age-marginalised SFH in which the largest stellar mass fraction is associated with an old population at $\sim 10$ Gyr, followed by a secondary peak at $\sim 4.8$ Gyr. More recently, analyses based on nearby \emph{Gaia} DR3 samples have revealed a still richer variety of solutions. \citet{mazzi2023} found an SFH that declines mildly between 13 and 3 Gyr, shows a pronounced enhancement between 3 and 1 Gyr ago, and is followed by a decline and a very recent increase ($\lesssim\!0.1$ Gyr). \citet{gallart2024} instead recovered an SFH with very low activity at the
oldest ages, followed by a sharp rise leading to a first major episode around
10~Gyr ago. Star formation then remains active, with multiple bursts superposed on a sustained level, until a marked dip at about 4~Gyr ago. It subsequently resumes at a higher average level and continues with a bursty behaviour to the present. Other groups have extended the analysis beyond the immediate solar neighbourhood to investigate the SFH on kiloparsec scales. For example, using \emph{Gaia} DR2 data, \citet{mor2019} found an SFH dominated by old populations, characterised by an overall decline from old to young ages and by an intermediate-age enhancement around 2-3 Gyr ago, whereas \citet{ruizlara2020} recovered a bursty SFH, with three conspicuous and relatively narrow episodes of enhanced star formation dated at 5.7, 1.9, and 1.0 Gyr ago.

A complementary approach comes from the white-dwarf (WD) population. The WD luminosity function (WDLF) provides an independent clock of the integrated history of star formation. ``WD cosmochronology'' dates back more than three decades \citep[e.g.][]{Winget1987,Liebert1988}, when the faint-end cutoff of the WDLF was recognised as a sensitive disk age indicator. Subsequent studies, combining improved cooling models with progressively larger local samples \citep[e.g.][]{Oswalt1996,Leggett1998,Harris2006,Giammichele2012}, generally placed the onset of disk star formation at $\sim8$-$11$~Gyr.

More recently, several studies have attempted to recover the detailed SFH
rather than only its onset time. \citet{Rowell2013} showed that the WDLF can be
inverted to recover a time-dependent SFH, finding a broadly bimodal solution
with enhanced activity at intermediate and old ages separated by a relative
lull. \citet{Tremblay2014} likewise recovered a broadly bimodal SFH, although
with different relative peak amplitudes.

Wider surveys and increasingly complete catalogues have progressively extended the reach of WD-based cosmochronology. Pan-STARRS1 pushed the WDLF to unprecedented depth \citep{Lam2019}, while \emph{Gaia} delivered homogeneous WD catalogues across the sky \citep{JimenezEsteban2018,GentileFusillo2019}. These developments enabled more detailed local reconstructions. { Using the spectroscopically confirmed 40 pc Gaia WD sample, \citet{Cukanovaite2023} found that the observed magnitude distribution is consistent with an approximately constant birth SFH\footnote{{ Here ``birth SFH'' denotes the intrinsic SFH before the effects of vertical dynamical heating in the local volume, in particular the age-dependent increase of the stellar scale height.}} over the past \(\sim 10.5\) Gyr.  This problem has recently been revisited by \citet{Roberts2025}, using the updated spectroscopically confirmed Gaia 40 pc WD census of \citet{Obrien2024}. Their direct-age reconstruction suggests an onset at \(10{-}12\) Gyr, relatively low activity at old and intermediate ages, and a rise toward a maximum at \(2{-}3\) Gyr. Using the \emph{Gaia} 100~pc WDLF, \citet{LamRowell2025} recovered a structured local SFH, with low activity at the oldest ages, enhanced star formation around 8--10~Gyr, reduced activity at intermediate ages, and a strong recent component.}

Overall, published reconstructions based on non-degenerate CMD fitting (hereafter CMD-derived SFH) and on WD cosmochronology do not yet converge on a unique local SFH. They differ in the relative importance of old, intermediate-age, and recent star formation, likely because of differences in stellar evolutionary tracks, sample selection, WD cooling models, and statistical methodology. At the same time, non-degenerate stars and WDs offer complementary views of the same underlying local SFH. At a fixed look-back time, CMD fitting is constrained by stars that are still present in non-degenerate evolutionary phases, whereas the WDLF samples, through the IFMR and the cooling sequences, the descendants of stars that have already left those phases.

In the present work, we first fit the \emph{Gaia} CMD of non-degenerate stars, from the pre-main sequence through the asymptotic giant branch, using the Star Formation Evolution Recovery Algorithm (\textsc{SFERA}) synthetic-population code \citep[e.g.][]{cignoni2015,Sacchi2018}. The fit is performed independently with two sets of stellar evolutionary models, providing two CMD-derived SFHs and age-metallicity relations (AMRs) for the local population. We then use each CMD-derived SFH to predict the corresponding local WD population: progenitor masses are drawn from the adopted IMF, pre-WD lifetimes are taken from the same stellar models used in the CMD analysis, final WD masses are assigned through an initial-final mass relation (IFMR), { WD atmospheric types are drawn according to the adopted DA/DB mixture}, and the resulting WDs are evolved along the corresponding cooling tracks to construct a synthetic WDLF. This WD-population calculation is repeated for four independent cooling grids. Comparing the full morphology of the resulting WDLFs with the
Gaia-quality-selected 40~pc WD sample provides an independent test of both
the CMD-derived SFHs and the adopted WD cooling-age scales.

To ensure statistical significance while minimising biases related to vertical stratification, we restrict our CMD fitting to a cylindrical volume of radius $200$~pc and vertical extent $|z|\leq40$~pc around the Sun. The $200$~pc radius yields a well-populated CMD, while the narrow vertical extent limits the impact of orbital diffusion and scale-height differences.\footnote{Adopting a larger vertical extent would mix populations with different kinematics and scale heights, making the comparison with the observed 40~pc WD sample less direct.} This choice allows us to rescale the recovered SFH to the $40$~pc volume for a direct comparison with the observed WDLF.

The paper is organised as follows. 
Section~2 describes the \textit{Gaia} sample selection. 
Section~3 presents the CMD-fitting technique, while Section~4 describes the treatment of observational uncertainties. 
Section~5 discusses the recovered SFH and AMR from the CMD analysis. 
Sections~6 and 7 describe the observed and synthetic WD samples, respectively. 
Section~8 presents the comparison between the observed and predicted WDLFs obtained with different cooling models, and Section~9 discusses the truncation test based on the Montr\'eal sequences. 
Finally, Section~10 summarises the main conclusions.

\section{Selecting \emph{Gaia} DR3 data}

The initial sample downloaded from the \emph{Gaia} DR3 archive includes all sources within a cylinder of radius 200 pc and vertical extent $|z|\leq40$ pc, corresponding to a total height of 80 pc, centred on the Sun. Objects with a relative parallax error greater than 0.1 ($\sigma_\pi/\pi > 0.1$) were removed from the data set. This filter ensures sufficiently accurate astrometric measurements and allows distances to be estimated from the inverse-parallax approximation (see, e.g., \citealt{Lindegren2021}). 
We further selected only sources with relative flux uncertainties better than 10\% in the $G$, $G_{BP}$, and $G_{RP}$ bands. This ensures that the transformation from fluxes to magnitudes remains in the high-S/N regime, where magnitude errors can be treated as approximately Gaussian (see, e.g., \citealt{riello2021}). { We also used the corrected BP/RP flux-excess factor \(C^*\)
defined by \citet{riello2021}, which measures the internal consistency of
the Gaia \(G\), \(G_{\rm BP}\), and \(G_{\rm RP}\) fluxes. We retained
sources satisfying
\[
|C^*| < 3\,\sigma_{C^*}(G),
\]
where \(\sigma_{C^*}(G)\) is the expected dispersion of \(C^*\) at the source \(G\) magnitude. No RUWE or IPD cuts were applied to this sample, to avoid rejecting unresolved binaries included statistically in the CMD modelling.}

\begin{figure*}
    \centering
    \includegraphics[width=0.95\linewidth]{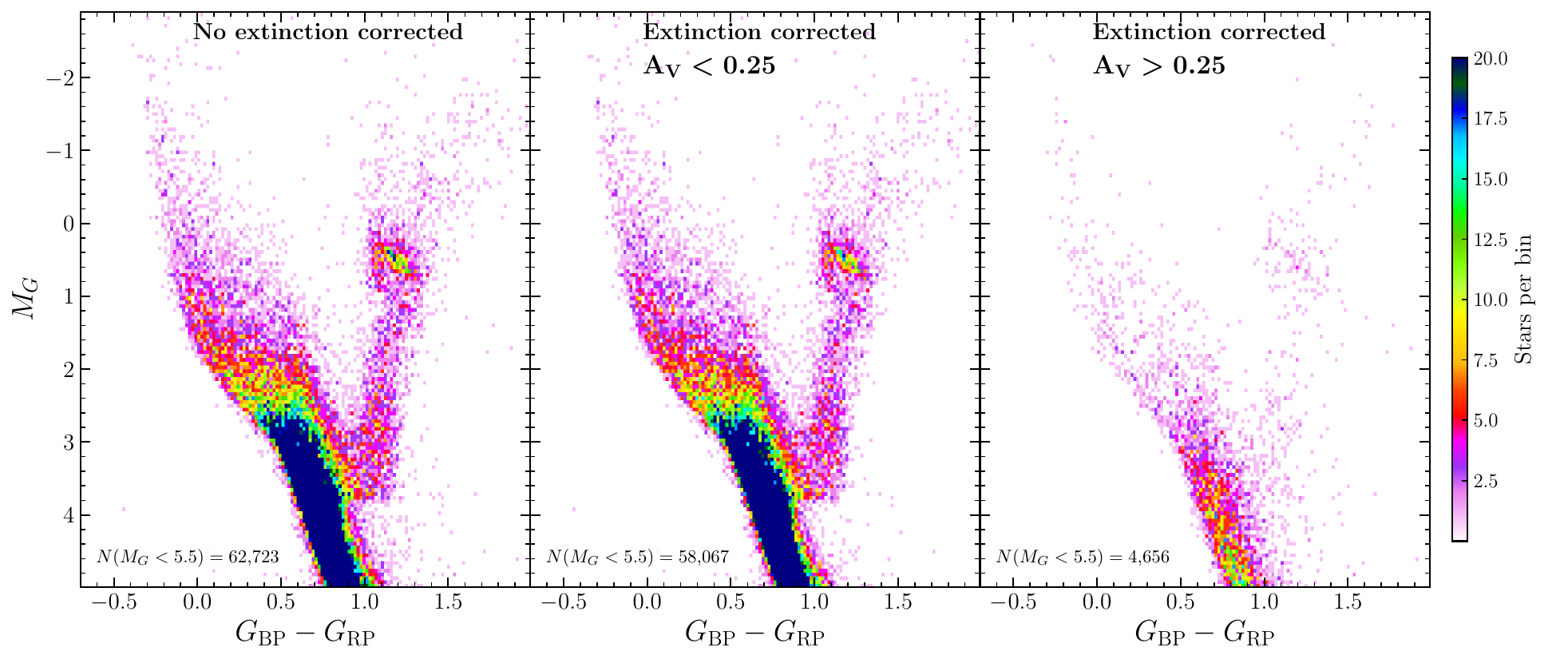}
\caption{CMD of stars in a cylindrical volume centred on the Sun, with
cylindrical radius up to 200~pc and vertical extent $|z|\leq40$~pc. Left
panel: CMD before extinction correction. Middle panel: CMD after extinction
correction, excluding objects with $A_V>0.25$. Right panel: CMD of the excluded
high-extinction subsample ($A_V>0.25$).}
    \label{3cmds}
\end{figure*}

Finally, we retained only stars brighter than $M_G = 5.5$ mag. This limit lies at least one magnitude below the oMSTO, which is located around $M_G \simeq 4$ mag, and therefore preserves the age-sensitive region of the CMD while excluding fainter stars that contribute little leverage on the recovered SFH. { After these astrometric, photometric, and magnitude cuts, the sample
contains 62,723 stars.} The resulting CMD is shown in the left panel of Fig.~\ref{3cmds}. 

Despite the proximity of our sample, signs of extinction are evident: several stars fall redward of the main red-giant branch and the lower main sequence. This shift cannot be attributed to photometric scatter alone, since the blue edge of the main sequence remains much sharper than its red boundary. Correcting for extinction is therefore essential in order to recover a reliable SFH. For this purpose, we adopted the recent 3D maps of \citet{vergely2022}, derived from the inversion of large spectroscopic and photometric surveys, including \emph{Gaia} DR3. We used the map covering a volume of $3 \times 3 \times 0.8$ kpc at 10 pc resolution, which is well matched to our nearby sample. Extinction values from the map are given in the \textit{V} band (central wavelength 551 nm) and were transformed into the \emph{Gaia} $G$, $G_{BP}$, and $G_{RP}$ passbands using the extinction coefficients provided in the \emph{Gaia} auxiliary data page\footnote{https://www.cosmos.esa.int/web/gaia/edr3-extinction-law}.

The central panel of Fig.~\ref{3cmds} presents the final CMD after applying extinction corrections and excluding stars with $A_V > 0.25$. The fact that the blue edge remains sharp supports the reliability of the adopted corrections, since large errors in the extinction estimates would otherwise broaden this boundary. The right panel shows only the high-extinction subsample ($A_V > 0.25$); its broadened morphology likely reflects large uncertainties in the extinction estimates, although a few objects could be genuine young pre-main-sequence stars. { The excluded high-extinction sources amount to only 8\% of the original sample, leaving 58,067 stars in the final CMD-fitting sample.}

\section{CMD fitting technique}

We reconstructed the SFH using the synthetic CMD-fitting technique
(e.g. \citealt{tosi1990,tolstoy1996,dolphin1997,dolphin2002,cignoni2006,
cignoni&tosi2010,aparicio&hidalgo2009,weisz2012,ruizlara2018,mor2019,
mazzi2023}) with the code SFERA.

SFERA represents the SFH as a weighted sum of $j\times k$ discrete
star-formation episodes, hereafter basis functions (BFs), each spanning the age
interval $[t_j,t_j+\Delta t]$ and the metallicity interval
$[Z_k,Z_k+\Delta Z]$. { To build each BF, we generate a large sample of synthetic stars by drawing
their ages uniformly from $[t_j,t_j+\Delta t]$, metallicities from
$[Z_k,Z_k+\Delta Z]$, and masses from a uniform distribution. Each star is
then weighted according to the \citet{kroupa2001} IMF. This sampling improves
the coverage of massive stars and short-lived evolutionary phases while
preserving the adopted IMF. Each BF contains approximately \(10^6\) synthetic
points, making finite-sampling fluctuations negligible.} In this
study, we adopt an age resolution $\Delta t$ of 50~Myr for $t \le 100$~Myr,
100~Myr for $100<t\le500$~Myr, 0.25~Gyr for $500$~Myr$<t\le2$~Gyr, 0.5~Gyr for
$2<t\le5$~Gyr, and 1~Gyr for epochs older than 5~Gyr. Metallicity is allowed to
vary over $[\mathrm{M/H}]=-1.5$ to $+0.5$ in steps of 0.1~dex. 

{ In this work we used two independent sets of stellar evolutionary models: the PARSEC library\footnote{Padova and Trieste Stellar Evolution Code v1.2S coupled with the COLIBRI AGB thermal-pulse code} \citep{Bressan2012,Tang2014,Marigo2017} and the PROSECCO/PISA models \citep{tognelli_cumulative_2015}. These libraries differ in their input physics and in the treatment of macroscopic processes such as convection, convective-boundary mixing, and mass loss. The synthetic CMD construction was performed separately for the two libraries, and their comparison is used to estimate the systematic uncertainty associated with the stellar models. Stellar magnitudes are then assigned
by interpolation on the adopted isochrone grid of the corresponding
stellar library\footnote{ Absolute Gaia
\(M_G\), \(G_{\rm BP}\), and \(G_{\rm RP}\) magnitudes are taken from
bolometric-correction tables based on the published Gaia EDR3 passbands.}.}

A fraction of the synthetic stars is paired with an unresolved binary companion, whose mass is likewise drawn from the same IMF and whose flux is summed with that of the primary. In this work, we assume an unresolved binary fraction of 30 per cent (see \citealt{cignoni2009} for an extensive discussion of this choice).

To produce realistic simulations, theoretical photometry is degraded to mimic the observational conditions of the data as described in Section \ref{err}. { Before constructing the Hess diagrams, both the observed and synthetic CMDs are restricted to the age-sensitive region extending to \(M_G\simeq4.3\). This limit includes the oMSTO over the full metallicity range of the stellar libraries, including the fainter turn-off of the most metal-rich models. The fitted region contains 32,179 observed stars.}

 Finally, BFs’ CMDs and the observational CMD are binned in n bins of colour and m bins of magnitude to construct density maps (Hess diagrams). The result is a library of $j \times k$  Hess diagrams  $BF_{m,n} (j, k)$, whose elements are linearly combined to match the observational counterpart, following the equation below:

\begin{equation}
\label{eq: bf_linear_comb}
    N_{n,m} = \sum_{j}\sum_{k} a(j,k)\times \text{BF}_{m,n}(j,k)  \,\,\, .
\end{equation}
The coefficients $a(j, k)$, representing the SFR at the time step $j$ and metallicity step $k$, are found by minimising a Poissonian likelihood (see e.g. \citealt{cash79}):

\begin{equation}
\label{eq: poisson_likely_sfera}
    \chi^{2}_{P} = 2 \sum_{i=1}^{N_{bin}} e_{i} - o_{i} + o_{i}\ln \frac{o_{i}}{e_{i}}  \,\,\,,
\end{equation}

where $e_i$ and $o_i$ are, respectively, the expected and the measured number of stars contained in the $i\text{th}$ cell of the binned CMDs. The uncertainties around the best-fit solution are estimated by bootstrapping the data and re-deriving the solutions.

From the minimisation we recover the coefficients $a_{j,k}$ which, together with the total mass $M_{j,k}$ and duration $\Delta t_j$ of each BF, define the most likely SFH behind the data. { Stars fainter than the fitted CMD region are not used as direct
constraints, but their contribution is included in the reported SFR by
integrating the adopted \citet{kroupa2001} IMF down to the
hydrogen-burning limit.} The mean metallicity at age $t_j$ is then computed as the mass‐weighted average of the metallicity bins $Z_k$:

\begin{equation}
Z(t_j)
= \frac{\displaystyle\sum_k Z_k \, a_{j,k}\,  M_{j,k}}
{\displaystyle\sum_k a_{j,k}\, M_{j,k}}
\end{equation}

Repeating this calculation at every age step $t_j$ yields an estimate of the AMR implied by the best‐fit model.

To explore the wide parameter space resulting from the choice of the age and metallicity steps, SFERA combines the genetic algorithm (GA) Pikaia\footnote{This is a public available routine developed at the High Altitude Observatory.} with a local search routine. { The GA explores the parameter space at many points simultaneously, making the solution less sensitive to initial conditions and local minima; the subsequent local search speeds up convergence and refines the best solution. Further details on this optimisation strategy are given in \citet{cignoni2015}.}

\section{Uncertainties}
\label{err}

The synthetic magnitudes obtained from the population-synthesis models are perturbed to account for observational uncertainties using an empirical, CMD-based error model. The observed \textit{Gaia} CMD, defined in the plane of absolute magnitude $M_G$ versus colour $(G_{BP}-G_{RP})$, is divided into cells whose size is chosen to ensure that each cell contains at least ten observed stars. This guarantees a statistically meaningful sampling of the local error distributions.

For each synthetic star, observational errors are assigned by sampling from the
distributions of \textit{Gaia} uncertainties of real stars located in the same
CMD cell. Specifically:

\begin{enumerate}
\item A photometric error in the $G$ band, $\Delta G$, is drawn from a Gaussian
distribution centred on zero, with a dispersion randomly selected from the
distribution of observed $\sigma_G$ values in the corresponding cell.
\item A parallax error, $\Delta \pi$, is drawn from a Gaussian distribution
centred on zero, with a standard deviation sampled from the observed
distribution of $\sigma_\pi$ in the same cell. Since the CMD fitting is
performed in absolute magnitude, this parallax perturbation is propagated into
an additional contribution to the absolute magnitude,
\begin{equation}
\Delta M_{G,\pi} \simeq \frac{5}{\ln 10}\,\frac{\Delta \pi}{\pi},
\end{equation}
which is valid given the imposed parallax-quality cuts. The total perturbation
applied to the synthetic absolute magnitude is therefore
\begin{equation}
\Delta M_G = \Delta G + \Delta M_{G,\pi}.
\end{equation}

\item Errors in the $G_{BP}$ and $G_{RP}$ bands, denoted as $\Delta G_{BP}$ and
$\Delta G_{RP}$, are drawn jointly from a bivariate Gaussian distribution in
order to preserve the empirically estimated correlation of the $G_{BP}$ and
$G_{RP}$ photometric uncertainties within each CMD cell. The covariance matrix
is estimated empirically within each CMD cell as
\begin{equation}
\Sigma =
\begin{pmatrix}
    \sigma^2_{G_{BP}} &
    \sigma_{G_{BP}}\sigma_{G_{RP}}\,\mathrm{corr}(G_{BP},G_{RP}) \\
    \sigma_{G_{BP}}\sigma_{G_{RP}}\,\mathrm{corr}(G_{BP},G_{RP}) &
    \sigma^2_{G_{RP}}
\end{pmatrix},
\end{equation}
where $\mathrm{corr}(G_{BP},G_{RP})$ is the correlation coefficient measured
from the observed stars in the cell. These perturbations are applied to the
intrinsic synthetic magnitudes, and the resulting colour error is given by
\begin{equation}
\Delta(G_{BP}-G_{RP}) = \Delta G_{BP} - \Delta G_{RP}.
\end{equation}
\end{enumerate}

This empirical approach allows the synthetic CMD to reproduce the magnitude and colour-dependent observational scatter of the \textit{Gaia} data using the uncertainties measured in the corresponding CMD cells, rather than imposing a global analytic error model.

\section{Recovered SFH}

\begin{figure*}[t]
    \centering
    \begin{minipage}[t]{0.74\textwidth}
        \vspace{0pt}
        \centering
        \includegraphics[width=\linewidth]{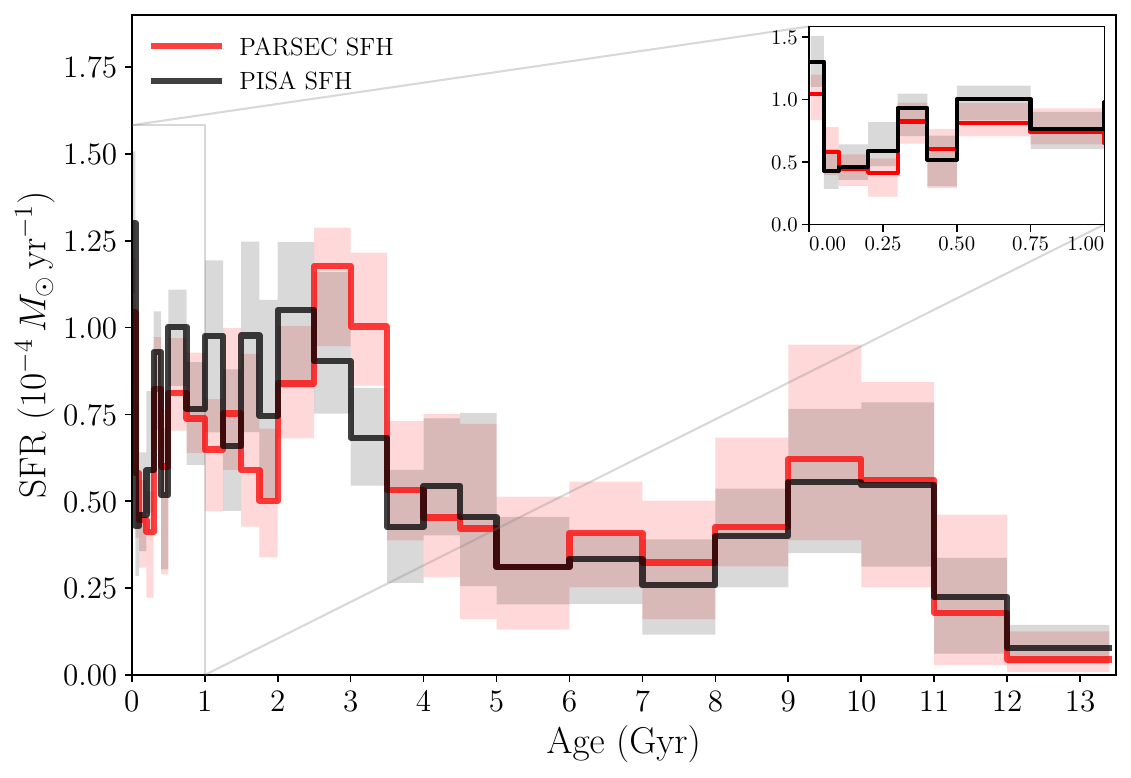}
    \end{minipage}
    \hfill
    \begin{minipage}[t]{0.23\textwidth}
        \vspace{0pt}
        \caption{Recovered SFHs as a function of look-back time for the
        PARSEC (red) and PISA (black) stellar models. The solid curves
        show the median solutions, and the shaded areas represent the
        corresponding uncertainty intervals. The inset provides a zoom
        into the most recent star-formation activity.}
        \label{sfh}
    \end{minipage}
\end{figure*}
  
Figure~\ref{sfh} shows the reconstructed SFH as a function of look-back time, using the two independent sets of stellar models. Bold lines trace the median solutions: the red curve corresponds to the PARSEC SFH and the black curve to the PISA SFH. The shaded areas mark the statistical uncertainties, defined as the 5th--95th percentile range of the SFHs obtained by re-fitting bootstrap realisations of the observed CMD, computed independently in each age bin\footnote{
Adjacent age bins should not be interpreted as fully independent.
Neighbouring BFs can populate partially overlapping regions of the CMD,
especially at old ages and in the presence of age-metallicity
degeneracy, so the fit can redistribute part of the SFR between nearby
age bins. The error bars shown for the SFH are marginal bootstrap
uncertainties and do not display these bin-to-bin correlations. For this
reason, we also consider the cumulative SFH, which is less sensitive to
local bin-to-bin anticorrelations.}.

At the earliest epochs (look-back time $>11\ \mathrm{Gyr}$), both SFH reconstructions predict an activity consistent with zero or only marginally above. They then rise to a first pronounced peak of $\sim0.7\times10^{-4}\,M_\odot\,\mathrm{yr}^{-1}$ between 9 and 11 Gyr ago. From 4 to 9 Gyr ago, the star-formation rate remains at a lower, nearly steady level of $\sim0.3$-$0.5\times10^{-4}\,M_\odot\,\mathrm{yr}^{-1}$. The most vigorous activity occurs between about 0.5 and 4 Gyr ago. In the PARSEC case, the rise is abrupt, reaching a sharp maximum of $\approx1.25\times10^{-4}\,M_\odot\,\mathrm{yr}^{-1}$ at 2.5 Gyr, whereas the PISA solution increases more gradually to a smoother peak of $\approx1.0\times10^{-4}\,M_\odot\,\mathrm{yr}^{-1}$ at 2 Gyr. Thus, PARSEC concentrates star formation into a shorter, more intense burst, while PISA distributes it over a longer interval. In the most recent half-billion years, both SFH reconstructions decline from their earlier peaks to a reduced rate of $\approx0.5\times10^{-4}\,M_\odot\,\mathrm{yr}^{-1}$. Overall, the two solutions share the same broad temporal pattern: an ancient low-activity phase, an early peak at 9-11 Gyr, a lower and relatively steady level between 4 and 9 Gyr, and a stronger episode of star formation at intermediate ages, culminating at $\sim2.5$ Gyr in the PARSEC solution and at $\sim2$ Gyr in the PISA one. Although the two models differ in the timing and sharpness of the main peak, they agree well on the overall rise-and-fall morphology and on the recent decline.

\begin{figure}
    \centering
 \includegraphics[width=0.95\linewidth]{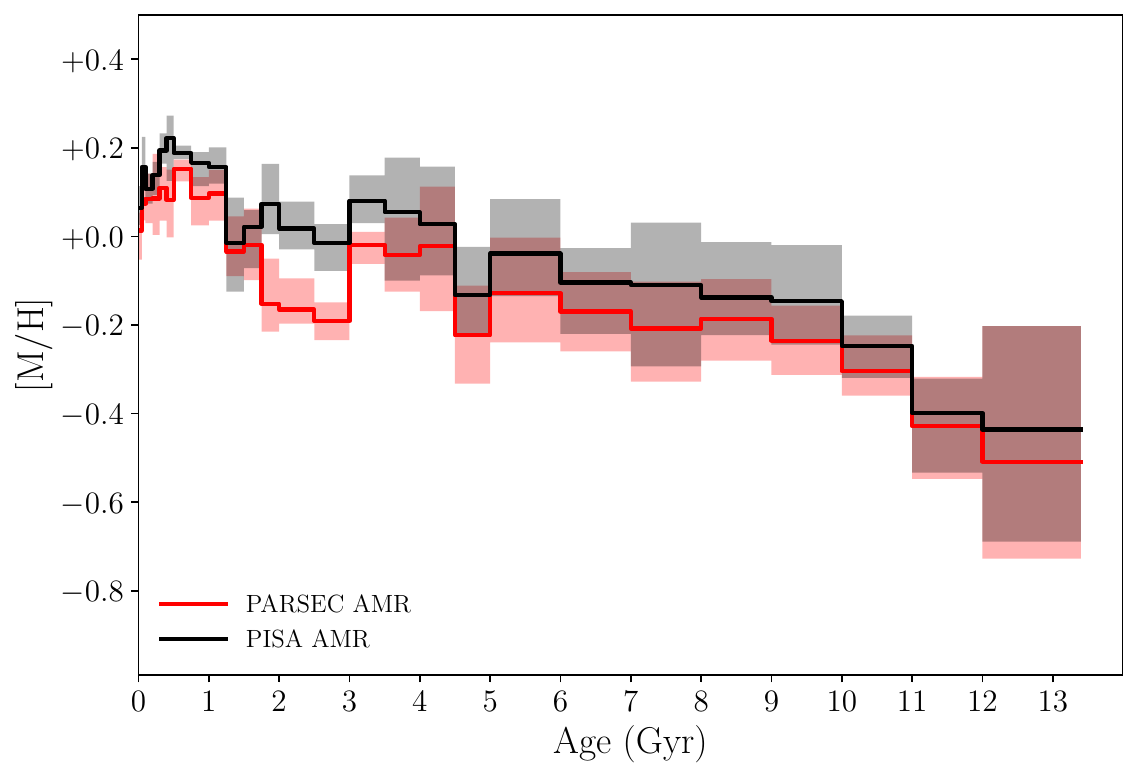}
    \caption{Recovered AMRs as a function of look-back time, obtained from CMD fitting with the PARSEC (red) and PISA (black) stellar models. Solid lines show the median relations, while the shaded regions indicate the corresponding uncertainty ranges. Both solutions display a global enrichment trend from old, metal-poor populations to younger, more metal-rich ones, with systematic differences between the two stellar libraries at intermediate ages.}
    \label{zt}
  \end{figure}

Figure~\ref{zt} presents the AMR recovered independently with the PARSEC (red) and PISA (black) stellar models. The vertical axis shows the recovered metallicity relative to the Sun, expressed as $[\mathrm{M/H}]\simeq\log(Z/Z_\odot)$ under the assumption of scaled-solar abundances, and the horizontal axis gives the look-back time. The shaded regions mark the 5th--95th percentile range of the corresponding bootstrap solutions in each age bin.

At look-back times beyond 11\,Gyr both AMRs lie at $\log(Z/Z_{\odot})\approx -0.5$\,dex, indicative of the earliest, metal-poor stellar populations. From roughly 11\,Gyr to 9\,Gyr ago the metallicity in both reconstructions rises gradually from $-0.5$\,dex to about $-0.2$ to $-0.1$\,dex. Between 9\,Gyr and 4\,Gyr ago the two curves remain approximately flat, offset by $\sim0.1$\,dex. In the interval 1-4\,Gyr ago the offset increases to $\sim0.2$\,dex, with the PISA solution systematically more metal-rich. Finally, during the last 0.5\,Gyr both tracks climb to their peak values of $\sim+0.2$\,dex.

\begin{figure}
    \centering
   \includegraphics[width=0.95\linewidth]{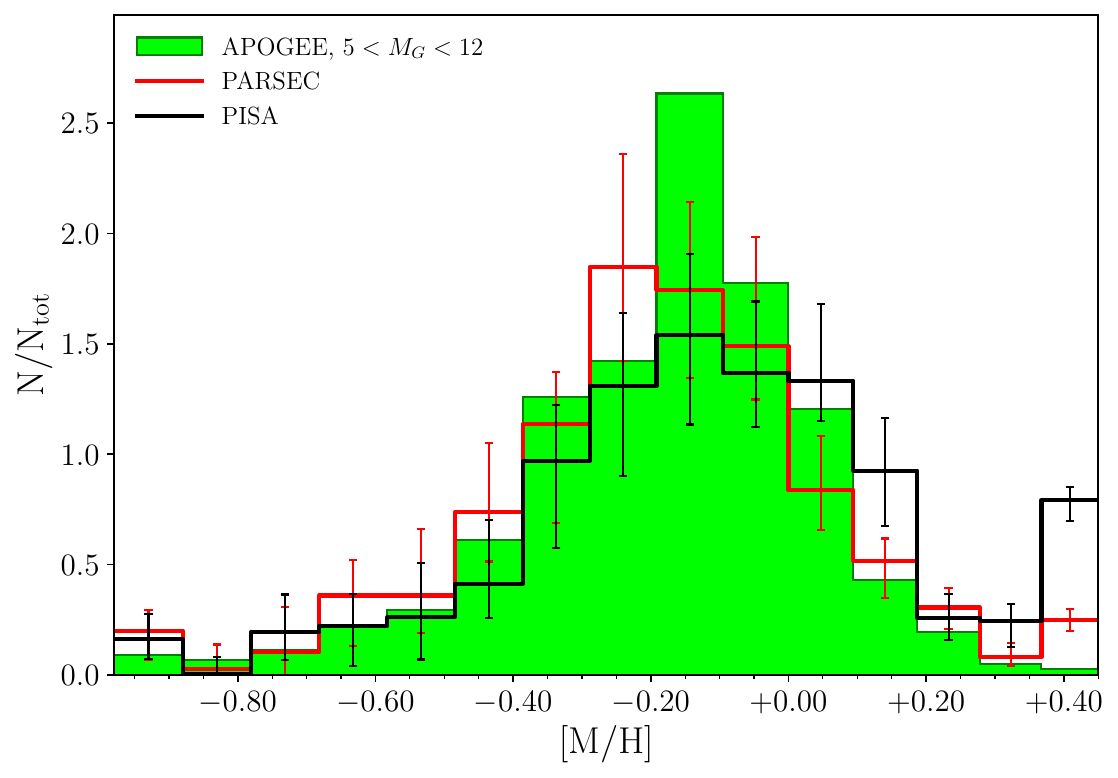}
    \caption{Comparison between the metallicity distributions predicted by the PARSEC (red) and PISA (black) solutions and the spectroscopic metallicity distribution observed by APOGEE (green histogram) in a similar local volume. The comparison provides an external check that the recovered SFH and AMR yield a chemically consistent description of the solar-neighbourhood population.}
    \label{nz}
\end{figure}

  In Fig.~\ref{nz}, we compare the metallicity distributions predicted by the PARSEC (red) and PISA (black) models with the spectroscopic metallicity distribution measured by APOGEE \citep{apogee2022} for stars within 40~pc (filled green histogram). { We use APOGEE stars satisfying \(5<M_G<12\), because restricting the comparison to the brighter CMD-fitting region, \(M_G\lesssim4.3\), would leave only a few RGB stars and no well-populated bright main sequence. The resulting metallicity distribution would therefore be dominated by small-number statistics and would not provide a representative sampling of the local age distribution.

The APOGEE distribution peaks at \([\mathrm{M/H}]\simeq-0.2\). Both synthetic distributions reproduce its broad location and width. PARSEC assigns somewhat more weight to the metal-poor side, whereas PISA produces a more extended super-solar tail. Given the limited size and non-trivial selection function of the local APOGEE sample, we regard this comparison as a qualitative consistency check rather than as an additional constraint on the fit.}

\begin{figure*}
    \centering
    \includegraphics[width=0.85\linewidth]{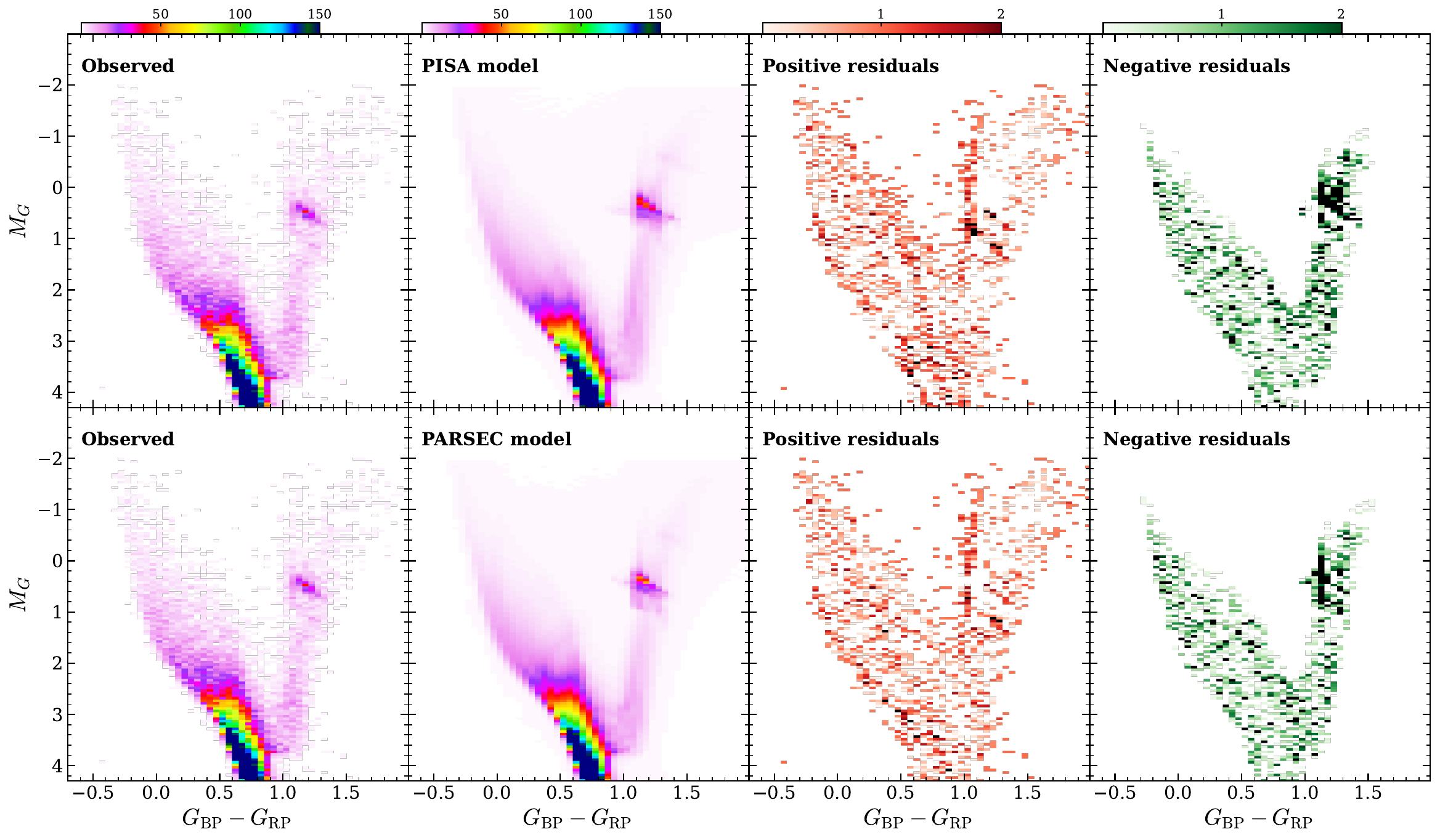}
    \caption{Comparison between the observed and synthetic \emph{Gaia} Hess
diagrams. Top row: PISA solution. Bottom row: PARSEC solution. In each row,
the panels show, from left to right, the observed Hess diagram, the synthetic
Hess diagram generated from the best-fitting SFH, the Poisson-normalised
residuals in bins where the observations exceed the model (red), and the
corresponding residuals in bins where the model exceeds the observations
(green).}
    \label{Residuals}
  \end{figure*}

Figure~\ref{Residuals} offers a direct comparison between the \emph{Gaia} CMD and our synthetic diagrams, using PISA models in the top row and PARSEC in the bottom. Each set of four panels, read from left to right, displays the observed Hess diagram, the synthetic Hess diagram derived from our best-fit SFH, the Poisson-normalised residuals $(O-M)/\sqrt{M}$ for bins where observations exceed the model (in red), and the corresponding residuals for bins where the model exceeds the data (in green).

Across most of the Hess diagram, our synthetic CMDs reproduce the \emph{Gaia}
Hess diagram very well, with 90\% of bins deviating by less than
2\,$\sigma$. The main coherent residual is located in the red-clump region,
which is populated by low-mass core-helium-burning stars. This offset is not
unexpected, since the colour and luminosity of red-clump stars depend
sensitively on age, metallicity, helium abundance, and the modelling of
core-helium-burning phases, and are not always reproduced quantitatively by
current isochrones \citep[e.g.][]{Plevne2020,Reyes2024,Chriss2025}. There, the
PARSEC model yields smaller residuals than PISA, reflecting a better match in
that evolutionary phase. In both model variants, the main sequence is
reproduced with high accuracy, supporting the robustness of the recovered SFH.

\begin{figure}
    \centering
    \includegraphics[width=0.9\linewidth]{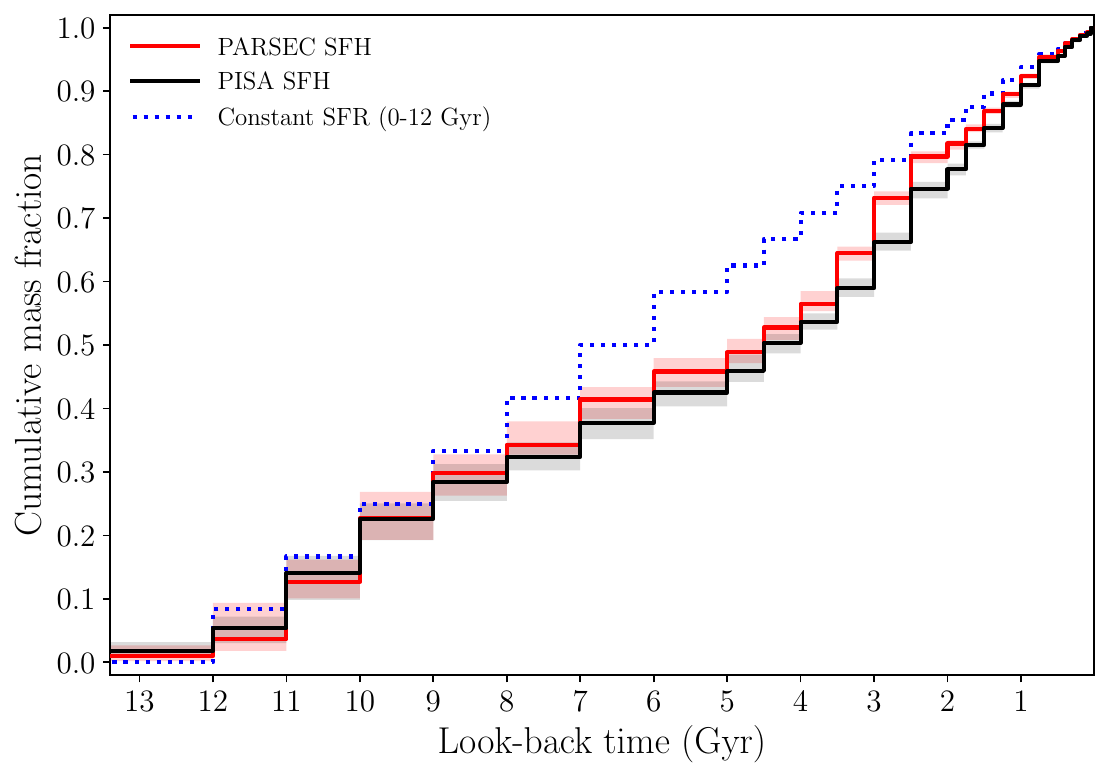}
    \caption{Cumulative mass fractions (CMFs) of the solar-neighbourhood disk derived from the CMD-based SFHs. Black and red curves show the PISA and PARSEC solutions, with shaded bands marking their uncertainties, while the blue dotted line corresponds to a constant star-formation rate over the past 12 Gyr.}
    \label{cum}
\end{figure}

To place our recovered SFH in the context of previous local studies, we compare the results in terms of the cumulative mass fraction (CMF), defined as the fraction of the total stellar mass formed up to a given look-back time. Compared to the differential SFH, the cumulative representation is less affected by correlations between adjacent time bins and less sensitive to differences in temporal binning, thus providing a cleaner basis for comparison across datasets. Figure~\ref{cum} shows the CMFs for the PISA (black) and PARSEC (red) solutions, together with the expected CMF for a constant star-formation rate over the last 12 Gyr (blue dotted line), shown here only as a reference case.

At early epochs ($\gtrsim 10$ Gyr) the PISA and PARSEC curves follow the constant-SFR reference closely. Divergences emerge at intermediate ages (3-8 Gyr), where both CMD-based solutions build up stellar mass more slowly than the constant case, indicating reduced activity during this period. In the most recent 2-3 Gyr the curves steepen, recovering the earlier deficit and converging toward the constant baseline by the present day. The two solutions remain very similar overall, with PARSEC rising slightly earlier and producing a somewhat narrower profile, while PISA yields a broader one. Taken together, the results point to a solar-neighbourhood disk that has grown steadily but not uniformly, with intermediate epochs contributing less to the cumulative mass than in the constant-SFR scenario.

Our cumulative SFHs (PISA and PARSEC) and that derived by \citet{gallart2024} from the \emph{Gaia} DR3 local 100~pc sample outline a broadly consistent scenario, but with measurable differences at early times.  At a look-back time of 10~Gyr our reconstruction reaches a mass fraction of about 0.15, whereas the Gallart et al. curves remain below 0.10. The half-mass point occurs between 4 and 5~Gyr in our reconstruction, and at slightly older ages, around 4.5--5~Gyr, in \citet{gallart2024}. At younger look-back times the agreement improves: the different
reconstructions reach a cumulative mass fraction of about 0.7 at roughly
2.5--3~Gyr and converge even more strongly at about 0.9, with roughly
90\% of the total mass already in place by 1--2~Gyr ago. Overall, compared to \citet{gallart2024}, our solution suggests
a somewhat larger early contribution and a slightly faster rise to the
half-mass point, while remaining in very good agreement at younger ages.

The SFHs derived by \citet{mazzi2023} and \citet{daltio2021} have a much
coarser temporal resolution at early epochs than our reconstruction, which
prevents a detailed assessment of the buildup at the oldest ages. Nevertheless,
the SFH of \citet{mazzi2023} for their innermost vertical slice,
$|z|<52.6$~pc, corresponding to the subsample closest to the Galactic plane and
therefore most comparable to our local volume, is broadly consistent with our
findings. It suggests that the cumulative mass fraction reaches $\sim0.5$ at a
look-back time of about 5~Gyr and $\sim0.9$ by about 1~Gyr ago. The case-C
solution of \citet{daltio2021}, based on a \emph{Gaia} DR2 local sample within
200~pc, is also broadly compatible, with a half-mass time of about
5.5--6~Gyr and $\sim0.9$ of the mass in place by about 1.5--2~Gyr ago.

The age--metallicity distribution inferred by \citet{alzate2021} for
\emph{Gaia} DR2 stars within 100~pc suggests a more old-weighted local history,
with stronger activity at early and intermediate ages, and therefore an earlier
mass buildup than in our reconstruction.

Among WD-based reconstructions, a qualitatively similar time sequence is also
recovered by the 100~pc WDLF inversion of \citet{LamRowell2025}. Their solution
shows very low activity at the oldest ages, followed by a broad old enhancement
around 8--10~Gyr. Star formation then decreases to a relative minimum at
intermediate ages, around 5--6~Gyr, before rising again into a stronger recent
episode over the last few Gyr, peaking around 3--4~Gyr. Although the two
analyses differ in sample volume, normalisation, and time resolution, this
agreement in broad temporal structure provides an additional consistency check
between our CMD-derived local SFH and WD-based reconstructions.

The SFHs discussed so far refer to the present-day stellar population contained within our thin local cylinder, and should therefore be interpreted as dynamically evolved local SFHs. In particular, the age dependence of the vertical scale height implies that stars of different ages are sampled with different efficiencies within the $|z|\leq 40$ pc slab: older populations are vertically hotter and more extended, so that a smaller fraction of their stars is expected to remain inside the adopted volume. Recovering an estimate closer to the intrinsic, vertically integrated SFH therefore requires an explicit assumption on the vertical density profile and on the age--scale-height relation.

\begin{figure}
    \centering
    \includegraphics[width=1\linewidth]{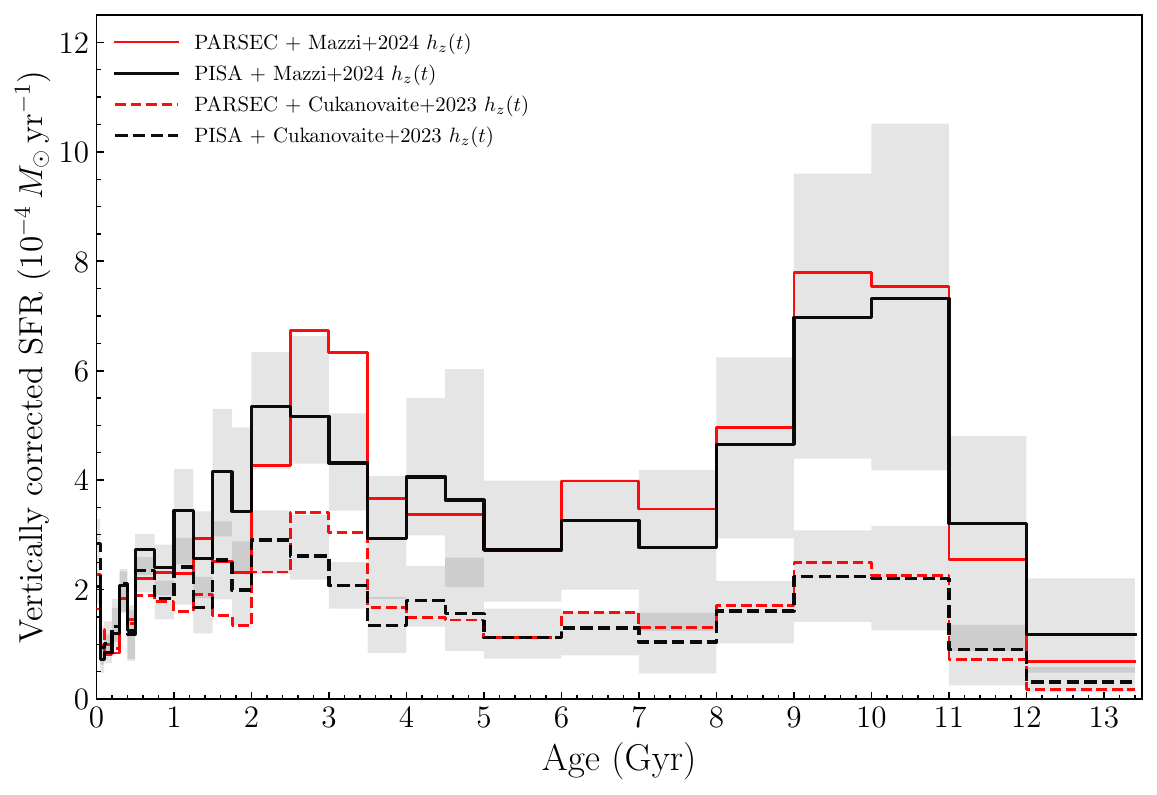}
\caption{Local SFHs corrected for the finite vertical extent of the observed
sample ($|z|\leq40$~pc), assuming an exponential vertical density profile and an
age-dependent scale height. Red and black curves show the PARSEC and PISA
solutions, respectively. Solid and dashed lines correspond to the
\citet{mazzi2023} and \citet{Cukanovaite2023} scale-height prescriptions,
respectively. Shaded regions show the bootstrap uncertainty ranges for the PISA
solutions only, included as representative uncertainties to avoid overcrowding
the figure.}
    \label{vertical}
  \end{figure}

Figure~\ref{vertical} illustrates the effect of this correction. We assume exponential vertical profiles and apply two independent age--scale-height prescriptions, from \citet{mazzi2023} and \citet{Cukanovaite2023}. Red and black curves show the PARSEC and PISA solutions, respectively, while solid and dashed lines correspond to the Mazzi et al. and Cukanovaite et al.  scale-height prescriptions, respectively. In both prescriptions, the correction becomes progressively stronger with increasing look-back time, because the fraction of stars sampled within the thin local slab decreases for older populations. { Although the size of the correction depends on the adopted prescription, both corrections enhance the early SFH relative to the dynamically evolved local history and produce a flatter reconstruction, approaching the nearly constant birth SFH inferred from the 40~pc WD sample by \citet{Cukanovaite2023}. The spread between the resulting histories provides an estimate of the systematic uncertainty associated with the adopted age--scale-height relation.}  
  
\section{WD sample}
To test the CMD-derived local SFH independently, we simulate the corresponding
WD population and compare it with the observed WD sample within 40~pc.

The observational sample was constructed by querying the \emph{Gaia}~DR3 archive\footnote{\url{https://gea.esac.esa.int/archive/}} and applying quality cuts aimed at ensuring reliable astrometry and photometry. As a first step, we selected all sources with parallaxes $\pi > 25$~mas, corresponding to distances within 40~pc, and with valid measurements in all three \emph{Gaia} photometric passbands.
To remove main-sequence contaminants and isolate the region of the CMD where WDs are expected to lie, we applied the broad cut defined by \citet{gentilefusillo2021}, \(M_G > 6 + (G_{BP} - G_{RP})\), retaining only objects located below this boundary.

We then rejected sources with relative uncertainties larger than 10\%
in parallax or in any of the three \emph{Gaia} photometric fluxes.
To further remove sources with unreliable photometric and astrometric
measurements, we used the prescription of \citet{fabricius2021},
requiring \(\mathrm{ipd\_frac\_multi\_peak}\leq2\),
\(\mathrm{ipd\_gof\_harmonic\_amplitude}<0.1\), and
\(\mathrm{RUWE}<1.4\), { and applied the same corrected BP/RP
flux-excess criterion used for the non-degenerate sample,
\(|C^*| < 3\,\sigma_{C^*}(G)\).}

{ After these selections, the final observed WD sample contains 988 objects
within 40 pc, fewer than the 1076 spectroscopically confirmed WDs in
\citet{Obrien2024} and the 1083 \citet{gentilefusillo2021} candidates. This is
expected, because our selection is designed to define a clean
Gaia-quality-selected WDLF rather than a complete 40 pc census.}

\section{WD synthetic population}

The construction of the synthetic WD population proceeds as follows. The CMD-derived SFH is recovered for a cylindrical volume of radius 200~pc and vertical extent $|z|\leq40$~pc. To compare it with the observed 40~pc WD sample, we first rescale the SFH from the CMD-fitting cylinder to a sphere of radius 40~pc. Since both samples have the same maximum vertical extent, $|z|\leq40$~pc, this rescaling is purely geometric.

We define $t_{\mathrm{up}}$ as the progenitor's evolutionary lifetime, i.e. the time elapsed from the pre-main sequence to the onset of the WD phase. The cooling time, $t_{\mathrm{cool}}$, is the time spent on the WD cooling track. The total age of a synthetic WD is therefore
\[
t_{\mathrm{tot}} = t_{\mathrm{up}} + t_{\mathrm{cool}} .
\]

In practice, the synthetic age sampled from the SFH corresponds to the total
stellar age, $t_{\mathrm{tot}}$. Once this age is assigned, we identify the
corresponding SFERA age bin and sample the progenitor metallicity from the mass
formed in its metallicity sub-bins.

Next, an initial mass $m_{\mathrm{ini}}$ is sampled from the adopted IMF. For this initial mass and progenitor metallicity, we determine from the stellar evolutionary models the time required to reach the end of the AGB phase, which we take as the onset of the WD phase and identify with the pre-WD lifetime $t_{\mathrm{up}}$.

Only synthetic stars satisfying
\[
t_{\mathrm{tot}} > t_{\mathrm{up}}
\]
are assumed to have evolved into WDs. Their cooling time is then given by
\[
t_{\mathrm{cool}} = t_{\mathrm{tot}} - t_{\mathrm{up}} .
\]
Stars with $t_{\mathrm{tot}}\leq t_{\mathrm{up}}$ have not yet reached the WD phase and are therefore discarded from the synthetic WD sample. { The baseline calculation assumes single-star evolutionary channels.  Unresolved non-interacting binaries are included only through their combined photometry, whereas interacting binary evolution and WD mergers are not modelled. Consequently, the very low-mass helium-core WDs produced through these channels are absent from the synthetic population.}

For the IFMR we adopt the piecewise linear relation of \citet{Cunningham2024}, derived from the \emph{Gaia} 40~pc WD sample, i.e. the same volume considered here. This choice provides a direct consistency with the observed WD population used for the WDLF comparison.

The WD mass, $m_{\mathrm{WD}}$, is obtained from $m_{\mathrm{ini}}$ by means of this IFMR. Absolute magnitudes $M_G$ and colours are then assigned to each synthetic WD from the cooling grid at the corresponding WD mass $m_{\mathrm{WD}}$, cooling time $t_{\mathrm{cool}}$, and atmospheric type. When the cooling grid includes a metallicity dependence, the progenitor metallicity is used to choose the corresponding cooling sequence.

Each synthetic WD must also be assigned a spectral type, since DA and DB WDs follow distinct cooling sequences. Atmospheric composition affects the cooling rate because hydrogen- and helium-dominated envelopes provide different outer boundary conditions and lead to different cooling timescales. Neglecting this distinction would therefore bias the predicted WDLF. To account for this, we randomly classify our synthetic WDs as DA or DB, following the cooling-track nomenclature and neglecting rarer spectral types, and adopt the observed DA fraction of $\sim72\%$ reported by \citet{esteban2023}.

We adopted four recent sets of WD cooling models: BaSTI\footnote{
For the BaSTI cooling models we used the version including the updated
conductive opacities of \citet{Blouin2020} in the H and He envelopes,
while retaining the \citet{Cassisi2007} conductive opacities in the C/O
core. These tracks also include \(^{22}{\rm Ne}\) diffusion in the
liquid phase.} \citep{salaris2022}, MIST \citep{bauer2026}, Montr\'eal \citep{bedard2020}, and La Plata \citep{althaus2013,camisassa2016,camisassa2017,camisassa2019}.  All these cooling grids provide the relation between cooling time and Gaia
magnitude, $t_{\rm cool}(M_G)$, at fixed WD mass and atmospheric type, but they
differ in their adopted input physics, core composition, crystallisation
treatment, and envelope structure. { In particular, La Plata includes
dedicated O/Ne-core sequences for massive WDs, whereas the Montr\'eal
and MIST grids assume C/O-core models over their tabulated mass ranges,
including the high-mass end. We do not attempt a process-by-process
attribution of the differences among the cooling grids, because the
published models differ simultaneously in several physical assumptions
and numerical implementations.}

Montr\'eal and La Plata models are computed
for a single reference metallicity, typically close to solar ($Z\simeq0.02$),
whereas BaSTI and MIST provide WD cooling sequences tabulated at several
metallicities. In our simulations, metallicity enters primarily through the progenitor
pre-WD lifetime. For BaSTI and MIST it also enters through the selection of the
appropriate cooling sequence, although this effect is secondary.

Using each cooling model set in turn, we generated synthetic WD populations and compared the resulting WDLFs in absolute $G$ magnitude with the observed \emph{Gaia} DR3 distribution within 40~pc.

\section{Results}

\begin{figure*}[h!]
    \centering
    \includegraphics[width=0.85\linewidth]{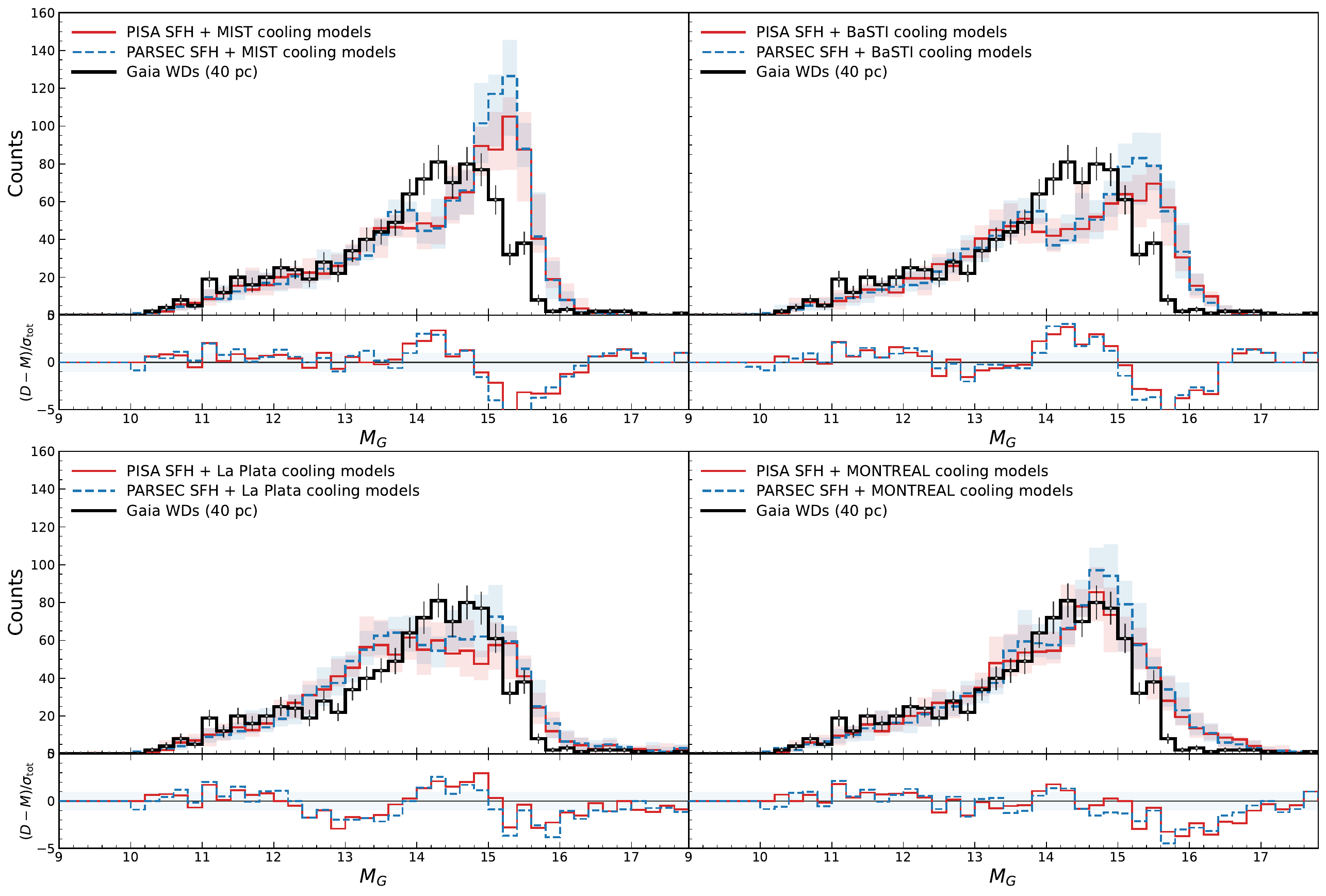}

    \caption{
    Observed \emph{Gaia} DR3 40~pc WDLF (black histogram) compared with
    synthetic predictions obtained from four independent WD cooling models:
    MIST (top left), BaSTI (top right), La Plata (bottom left), and
    Montr\'eal (bottom right). In each panel, synthetic WDLFs are computed
    using the same IFMR \citep{Cunningham2024} and the two CMD-based
    star-formation histories: PISA (red solid line) and PARSEC
    (blue dashed line). Shaded regions indicate the 5--95th percentile
    range of the synthetic realisations. Lower sub-panels show the residuals
    between the observed ($D$) and synthetic ($M$) WDLFs, computed as
    $(D-M)/\sigma_{\rm tot}$, where
    $\sigma_{\rm tot}=\sqrt{\sigma_D^2+\sigma_M^2}$ is the total uncertainty
    obtained by adding in quadrature the observational and model
    uncertainties.
    }

    \label{WD_LFs}
  \end{figure*}
  
  Figure~\ref{WD_LFs} compares the observed \emph{Gaia} DR3 40~pc WDLF (black histogram) with synthetic predictions obtained using the four independent sets of WD cooling models (MIST, BaSTI, La Plata, and Montr\'eal), a common IFMR \citep{Cunningham2024}, and the two CMD-based SFHs (PISA in red, PARSEC in blue). For each cooling-model set, the synthetic WDLFs are compared directly with the observations, and the corresponding residuals are shown in the lower panels.

In terms of look-back time $t$, the WDLF per unit magnitude can be written as
follows \citep[e.g.][]{fontaine2001}:
\begin{equation}
\begin{aligned}
\Phi(M_G) &=
\int_{0}^{T}\mathrm{d}t 
\int \mathrm{d}Z
\int_{M_{\min}}^{M_{\max}}\mathrm{d}M_{\rm ini}\;
\psi(t,Z)\,\phi(M_{\rm ini}) \\
&\quad \times
\left|
\frac{\partial t_{\rm cool}}{\partial M_G}
\right|
\delta\!\left[
t - t_{\rm up}(M_{\rm ini},Z) - t_{\rm cool}
\right] .
\end{aligned}
\end{equation}

Here $t$ is the look-back time, $T$ is { the time since the onset of star formation}, $\psi(t,Z)$ is the star-formation-rate distribution in the age--metallicity plane recovered from the CMD, and $\phi(M_{\rm ini})$ is the IMF. The SFH shown in the previous sections corresponds to the metallicity-integrated form of $\psi(t,Z)$. The pre-WD lifetime $t_{\rm up}$ is set by the initial mass and metallicity of the progenitor, whereas the cooling time $t_{\rm cool}$ is set by the WD mass, atmospheric type, chemical composition, and adopted cooling prescription. The factor $\left|\partial t_{\rm cool}/\partial M_G\right|$ is the residence time per magnitude interval along the cooling sequence.

This expression highlights the ingredients that determine the WDLF morphology shown in Fig.~\ref{WD_LFs}. The cooling prescription affects the cooling-age scale and the residence-time term, and therefore affects how WDs are mapped into magnitude once they have formed.  Differences between the PISA and PARSEC solutions enter instead through the recovered $\psi(t,Z)$ and the associated pre-WD lifetimes, which set the timing at which progenitors become WDs. The resulting WDLF morphology is thus set by the combined effect of WD production and subsequent cooling, with the latter becoming increasingly dominant along the faint tail.

Across all cooling prescriptions, the synthetic WDLFs reproduce the broad rise
of the observed distribution along the bright and intermediate branches. In the
synthetic populations, WDs in the range $10\lesssim M_G\lesssim13$ are mostly
recent arrivals on the cooling sequence, with $t_{\rm cool}\lesssim1$~Gyr,
$t_{\rm tot}\lesssim3$~Gyr, and a median
$t_{\rm cool}/t_{\rm tot}\simeq0.3$. This part of the WDLF therefore mainly
traces the recent WD production rate, set by the CMD-derived SFH, the IFMR, and
the pre-WD lifetimes. The cooling prescription affects the mapping into
$M_G$, but its impact is secondary here compared with the maximum and faint
tail.

The comparison becomes more discriminating toward the maximum and along the
post-maximum decline, where the different cooling prescriptions produce
distinct residual patterns.

Within each cooling prescription, differences between the two CMD-based SFHs
are also visible, although they remain smaller than the differences among
cooling grids. Their effect is most apparent around the WDLF maximum. In the
synthetic populations, the maximum region
($14.0\lesssim M_G\lesssim15.0$) is populated mainly by WDs with total ages
$t_{\rm tot}\sim5$--10~Gyr and cooling times
$t_{\rm cool}\sim3$--6~Gyr. The ratio $t_{\rm cool}/t_{\rm tot}$ spans a broad
range but is weighted toward high values, indicating that the maximum is
already substantially affected by WD cooling while still retaining sensitivity
to the timing of WD production. Its morphology is therefore shaped by the
combined effect of the CMD-derived SFH, pre-WD lifetimes, the cooling-age
scale, and the residence time along the cooling sequence.

Along the faint tail ($M_G\gtrsim15.0$), the synthetic WDs are instead strongly cooling-dominated, with $t_{\rm cool}/t_{\rm tot}$ concentrated toward high values. The faint tail is therefore especially sensitive to the cooling-age scale, while the old SFH component sets the number of old WDs that can populate this region.

Having described the regimes sampled by the different parts of the WDLF, we now examine the residual patterns produced by each cooling model set.

The MIST cooling models (top left panel) reproduce the bright and intermediate
branches comparably well, but place the post-maximum decline at systematically
fainter magnitudes and produce a maximum that is narrower than observed. As a
consequence, the synthetic WDLF shows a deficit around the observed maximum and
a clear excess at $M_G \gtrsim 15$. This behaviour is directly reflected in the
residuals, which display coherent positive values around the maximum and
negative residuals along the faint tail, indicating that, for the adopted SFHs,
the MIST sequences place too much weight at magnitudes fainter than the
observed maximum.

The BaSTI models (top right panel) show a similar qualitative behaviour. They reproduce the rise up to $M_G \approx 13.5$, but then develop an extended plateau between $M_G \approx 13.5$ and 14.5 that has no counterpart in the data. The maximum and the subsequent decline are shifted toward fainter magnitudes, again leading to an excess of objects beyond $M_G \sim 15.5$. The residual maps highlight this mismatch, with systematic positive residuals around the observed maximum and negative residuals along the faint tail.

The La Plata cooling models (bottom left panel) shift the post-maximum decline toward brighter magnitudes and substantially reduce the excess of objects at faint magnitudes, bringing the synthetic faint tail into closer agreement with the \emph{Gaia} histogram than in the MIST and BaSTI cases. This improvement, however, is not uniform over the full WDLF: the models show larger residuals at intermediate magnitudes, especially around $M_G\simeq12.5$--13.5, and still do not fully reproduce the height and shape of the observed maximum. Thus, La Plata provides a better match to the faint tail, but not to the full WDLF morphology.

Finally, the Montr\'eal models (bottom right panel) give the most balanced
agreement with the observed WDLF. They reproduce the bright and intermediate
branches and place the maximum and post-maximum decline close to the observed
positions. At fainter magnitudes ($M_G \gtrsim 15.5$), however, the residuals
increase again, indicating that the faint tail remains less well matched than
with the La Plata models, although the discrepancy is smaller than for MIST and
BaSTI.

Overall, the different cooling prescriptions leave clearly distinct signatures in the residuals. MIST and BaSTI shift too much weight to magnitudes fainter than the observed maximum, producing synthetic WDLFs whose maxima and post-maximum declines occur at too faint magnitudes and remain too populated along the faint tail. La Plata largely removes this faint-end excess but leaves residual discrepancies in the shape of the maximum. Montr\'eal provides the best compromise, reproducing the observed maximum region most accurately while still showing some residual excess at the faintest magnitudes.

\section{Total-age cut test}
\label{sec:agecut}

\begin{figure}[h!]
   \centering
    \begin{minipage}{0.85\linewidth}
        \centering
        \includegraphics[width=\linewidth]{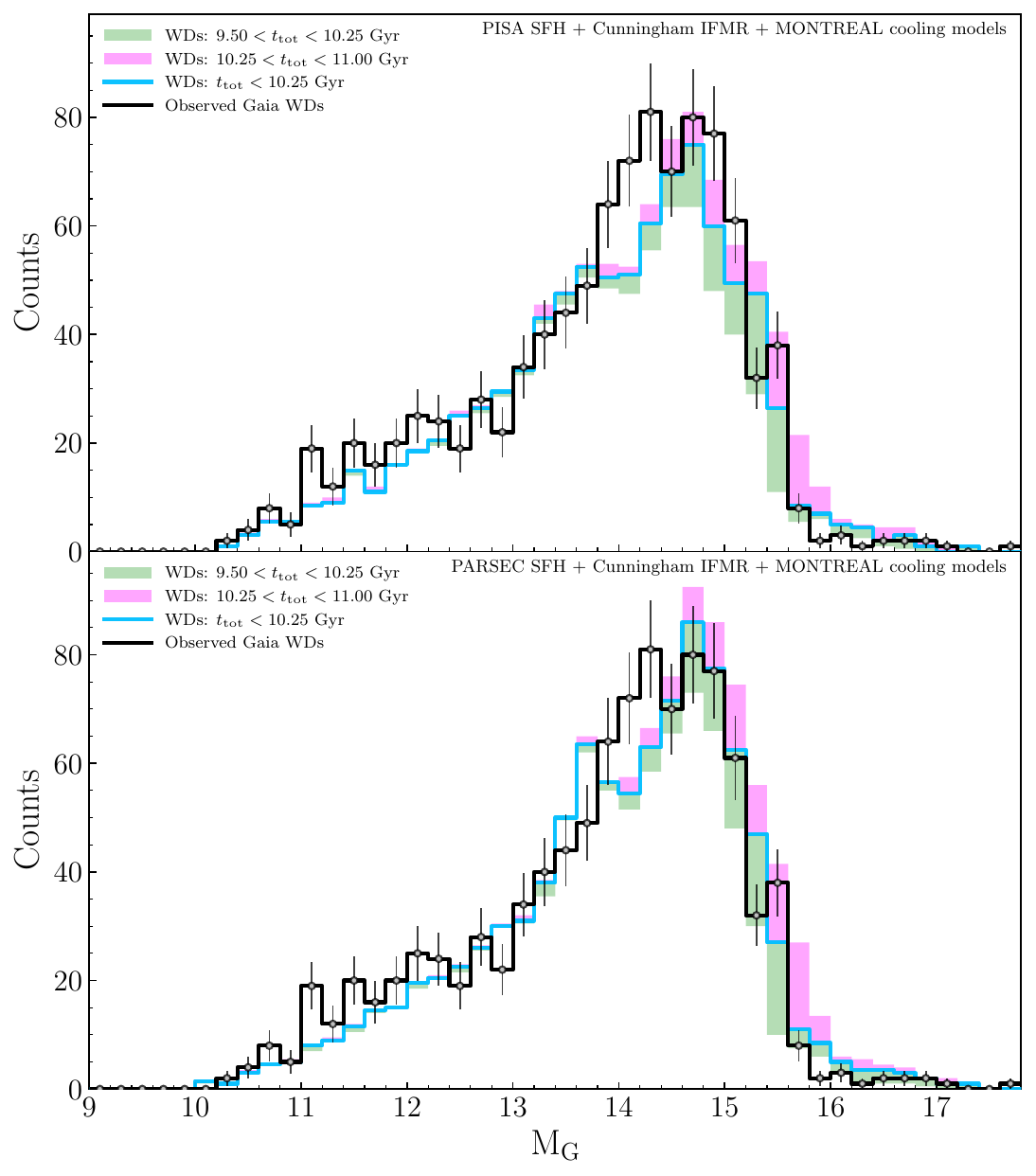}
    \end{minipage}
 \caption{Observed \textit{Gaia} DR3 40~pc WDLF (black histogram) compared
with synthetic WDLFs obtained from the PISA (top panel) and PARSEC (bottom
panel) CMD-based SFHs. The coloured envelopes show the effect of progressively
removing the oldest WDs from the synthetic population. The green envelope is
bounded by the WDLFs obtained with $t_{\rm tot}<9.50$~Gyr and
$t_{\rm tot}<10.25$~Gyr, while the magenta envelope is bounded by the WDLFs
obtained with $t_{\rm tot}<10.25$~Gyr and $t_{\rm tot}<11.00$~Gyr.}
    \label{WD_LFs_cut}
  \end{figure}

Figure~\ref{WD_LFs_cut} compares the observed \textit{Gaia} DR3 40~pc WDLF
with age-cut synthetic WDLFs obtained from the PISA and PARSEC CMD-based SFHs.
The upper and lower panels refer to the PISA and PARSEC solutions, respectively.
The test is performed with the Montr\'eal cooling sequences, which provide the
best match to the observed maximum and post-maximum decline in
Fig.~\ref{WD_LFs}.

The coloured envelopes show how the WDLF changes when upper cuts are applied to
the total WD age, thereby progressively removing the oldest WDs from the
synthetic population. The magenta envelope brackets the difference between the
cuts $t_{\rm tot}<11.00$~Gyr and $t_{\rm tot}<10.25$~Gyr, while the green
envelope brackets the difference between $t_{\rm tot}<10.25$~Gyr and
$t_{\rm tot}<9.50$~Gyr.

These cuts mainly affect the region beyond the WDLF maximum, especially the
post-maximum decline and the faint tail. Since WDs in this magnitude range are
cooling-dominated, lowering the upper limit on $t_{\rm tot}$ progressively
removes the oldest, longest-cooling objects from the synthetic population.
Within the adopted Montr\'eal cooling scale, the observed faint tail is better
reproduced when the synthetic population is truncated at a maximum total age of
about 11~Gyr. We therefore interpret this value as the maximum age of the local
Galactic disk implied by the WDLF, conditional on the adopted CMD-derived SFH,
IFMR, DA/DB mixture, and WD cooling prescription.

This truncation test is diagnostic and does not replace the CMD best-fitting
SFHs. The stellar mass involved is small: from the cumulative SFHs in
Fig.~\ref{cum}, only about 4--5\% of the local stellar mass is formed at
look-back times older than 11~Gyr. Removing this component has little effect on
the global CMD-derived SFH, but it affects the faintest WDs, which selectively
trace the oldest, longest-cooling objects.

\section{Conclusions}

This work uses WDs as an independent chronometer to test two CMD-derived
reconstructions of the solar-neighbourhood SFH. The non-degenerate CMD and the
WDLF probe the same underlying history in complementary ways. At a fixed
look-back time, the CMD fitting is constrained by stars that are still present
in non-degenerate evolutionary phases, while the WDLF samples, through the IFMR
and the cooling sequences, the descendants of stars that have already left those
phases. Their comparison therefore provides an independent consistency check on
both the recovered SFHs and the adopted WD cooling models.

We derived the SFH and AMR of the solar neighbourhood by fitting the
\emph{Gaia} DR3 CMD within a local cylinder of radius 200~pc and vertical extent
$\pm40$~pc, using the PISA and PARSEC stellar evolutionary models. From each
best-fitting CMD solution we generated synthetic WD populations adopting a
common IFMR \citep{Cunningham2024} and four cooling model sets: BaSTI, MIST, La
Plata, and Montr\'eal. The resulting synthetic WDLFs were compared with the
observed \emph{Gaia} DR3 40~pc WDLF.

The main conclusions are the following.

\begin{itemize}

\item The two CMD-based SFHs recover the same broad temporal behaviour: low or marginal activity at the oldest ages, an early episode around 9--11~Gyr ago, a lower and relatively steady level between about 4 and 9~Gyr, and enhanced star formation over the last few Gyr. The PISA and PARSEC solutions differ in the sharpness and timing of the intermediate-age enhancement, but agree on the overall mass build-up. In cumulative form, our reconstruction is broadly compatible with recent
Gaia-based local SFHs. The comparison with \citet{gallart2024}, which has the
highest temporal resolution among these recent studies, shows a similar overall
mass build-up, although our solution suggests a slightly larger early
contribution.

\item The recovered AMRs provide a chemically consistent description of the local population. The metallicity distributions predicted from the two CMD-based solutions reproduce the main APOGEE metallicity peak and the overall metal-poor tail. { Remaining differences are mostly confined to the relative weight of the metal-poor and super-solar tails. }

\item { Once local and birth SFHs are distinguished, our results are broadly consistent with WD-based reconstructions. Our uncorrected local SFHs show the same broad old enhancement, intermediate-age minimum, and stronger recent activity found by \citet{LamRowell2025}. A similar rise toward a \(2{-}3\)~Gyr maximum is found in the direct-age distribution of \citet{Roberts2025} after correcting for stars that have not yet evolved into WDs, but before applying their kinematic correction. Applying age-dependent vertical scale-height corrections up-weights old populations and produces flatter histories, bringing them closer to the nearly constant birth SFH inferred by \citet{Cukanovaite2023}.}

\item The bright and intermediate portions of the WDLF are reproduced reasonably
well for both CMD-based SFHs and for all cooling grids. In the synthetic
populations, WDs with $10\lesssim M_G\lesssim13$ have short cooling times and are dominated
by young total ages, with typical $t_{\rm tot}$ values below $\sim3$~Gyr and a
median $t_{\rm cool}/t_{\rm tot}$ of about 0.3. This magnitude range therefore
mainly probes the recent WD production rate, and hence the young part of the
CMD-derived SFH.

\item The WDLF maximum samples an intermediate regime in which both the timing of
WD production and the subsequent cooling evolution matter. In the synthetic
populations, this region is dominated by WDs with
$t_{\rm tot}\sim5$--10~Gyr and $t_{\rm cool}\sim3$--6~Gyr. Its height and width
therefore depend on the intermediate-age CMD-derived SFH, the AMR-dependent
pre-WD lifetimes, and the cooling time. The PISA--PARSEC differences
are visible in this region, but remain smaller than the systematic differences
among WD cooling grids.

\item The post-maximum decline and the faint tail are primarily controlled by the WD cooling prescription. MIST and BaSTI place too much weight at magnitudes fainter than the observed maximum and overpopulate the faint tail. La Plata substantially reduces this faint-end excess and gives the best match to the faint tail, but leaves larger residuals at intermediate magnitudes and around the maximum. Montr\'eal provides the best overall compromise, reproducing the observed maximum and post-maximum decline most accurately, while still leaving some residual discrepancies at the faintest magnitudes. { The WDLF comparison is therefore limited by the WD cooling models themselves: the same CMD-derived SFH produces substantially different WDLFs depending on the adopted cooling grid.}

\item { Since the Montr\'eal cooling models provide the best overall agreement with the observed WDLF,} we tested how the WDLF changes when the oldest WDs are progressively removed from the synthetic population. These cuts mainly affect the cooling-dominated post-maximum decline and faint tail. Within the adopted IFMR, CMD-derived SFH, and Montr\'eal cooling scale, the observed faint tail disfavours a substantial contribution from WDs with total ages older than 11~Gyr. This is broadly consistent with the maximum Galactic-disk age of $10.6\pm0.5$~Gyr inferred by \citet{Cukanovaite2023} from the \emph{Gaia} 40~pc WD sample.

\end{itemize}

{ A natural extension of this work is a joint fit in which non-degenerate
stars and WDs are modelled simultaneously, rather than using the WDLF only
as an external consistency check. Such an approach would exploit their
complementary age sensitivity and constrain the SFH, IFMR, DA/DB fraction,
and WD cooling parameters within a single population model.}

\begin{acknowledgements}
MC, PGPM, and SD acknowledge the INFN iniziativa specifica TAsP.
{ We thank the anonymous referee for carefully reading the manuscript and
for constructive comments that improved it. This work has made use of data
from the ESA mission \emph{Gaia}, processed by the Gaia Data Processing and
Analysis Consortium (DPAC).}
\end{acknowledgements}

\makeatletter
\def\bibcite#1#2{%
  \@ifundefined{b@#1}{%
    \global\@namedef{b@#1}{#2}%
  }{}%
}
\makeatother

\bibliographystyle{aa}
\bibliography{biblio}

@ARTICLE{soderblom2010,
       author = {{Soderblom}, David R.},
        title = "{The Ages of Stars}",
      journal = {\araa},
         year = 2010,
        month = sep,
       volume = {48},
        pages = {581-629},
          doi = {10.1146/annurev-astro-081309-130806},
archivePrefix = {arXiv},
       eprint = {1003.6074},
 primaryClass = {astro-ph.SR},
       adsurl = {https://ui.adsabs.harvard.edu/abs/2010ARA&A..48..581S}
}

@INPROCEEDINGS{soderblom2015,
       author = {{Soderblom}, David R.},
        title = "{Ages of Stars: Methods and Uncertainties}",
    booktitle = {Asteroseismology of Stellar Populations in the Milky Way},
         year = 2015,
       editor = {{Miglio}, Andrea and {Eggenberger}, Patrick and {Girardi}, L{\'e}o and {Montalb{\'a}n}, Josefina},
       series = {Astrophysics and Space Science Proceedings},
       volume = {39},
        month = jan,
        pages = {3},
          doi = {10.1007/978-3-319-10993-0_1},
archivePrefix = {arXiv},
       eprint = {1409.2266},
 primaryClass = {astro-ph.SR},
       adsurl = {https://ui.adsabs.harvard.edu/abs/2015ASSP...39....3S}
}

@book{Hipparcos1997,
  author    = {{ESA}},
  title     = {The Hipparcos and Tycho Catalogues},
  year      = {1997},
  publisher = {ESA Publications Division},
  series    = {ESA SP},
  volume    = {1200},
  address   = {Noordwijk, The Netherlands}
}

@ARTICLE{GaiaDR3,
       author = {{Gaia Collaboration} and {Vallenari}, A. and {Brown}, A.~G.~A. and {Prusti}, T. and {de Bruijne}, J.~H.~J. and {Arenou}, F. and {Babusiaux}, C. and {Biermann}, M. and {Creevey}, O.~L. and {Ducourant}, C. and {Evans}, D.~W. and {Eyer}, L. and {Guerra}, R. and {Hutton}, A. and {Jordi}, C. and {Klioner}, S.~A. and {Lammers}, U.~L. and {Lindegren}, L. and {Luri}, X. and {Mignard}, F. and {Panem}, C. and {Pourbaix}, D. and {Randich}, S. and {Sartoretti}, P. and {Soubiran}, C. and {Tanga}, P. and {Walton}, N.~A. and {Bailer-Jones}, C.~A.~L. and {Bastian}, U. and {Drimmel}, R. and {Jansen}, F. and {Katz}, D. and {Lattanzi}, M.~G. and {van Leeuwen}, F. and {Bakker}, J. and {Cacciari}, C. and {Casta{\~n}eda}, J. and {De Angeli}, F. and {Fabricius}, C. and {Fouesneau}, M. and {Fr{\'e}mat}, Y. and {Galluccio}, L. and {Guerrier}, A. and {Heiter}, U. and {Masana}, E. and {Messineo}, R. and {Mowlavi}, N. and {Nicolas}, C. and {Nienartowicz}, K. and {Pailler}, F. and {Panuzzo}, P. and {Riclet}, F. and {Roux}, W. and {Seabroke}, G.~M. and {Sordo}, R. and {Th{\'e}venin}, F. and {Gracia-Abril}, G. and {Portell}, J. and {Teyssier}, D. and {Altmann}, M. and {Andrae}, R. and {Audard}, M. and {Bellas-Velidis}, I. and {Benson}, K. and {Berthier}, J. and {Blomme}, R. and {Burgess}, P.~W. and {Busonero}, D. and {Busso}, G. and {C{\'a}novas}, H. and {Carry}, B. and {Cellino}, A. and {Cheek}, N. and {Clementini}, G. and {Damerdji}, Y. and {Davidson}, M. and {de Teodoro}, P. and {Nu{\~n}ez Campos}, M. and {Delchambre}, L. and {Dell'Oro}, A. and {Esquej}, P. and {Fern{\'a}ndez-Hern{\'a}ndez}, J. and {Fraile}, E. and {Garabato}, D. and {Garc{\'\i}a-Lario}, P. and {Gosset}, E. and {Haigron}, R. and {Halbwachs}, J.-L. and {Hambly}, N.~C. and {Harrison}, D.~L. and {Hern{\'a}ndez}, J. and {Hestroffer}, D. and {Hodgkin}, S.~T. and {Holl}, B. and {Jan{\ss}en}, K. and {Jevardat de Fombelle}, G. and {Jordan}, S. and {Krone-Martins}, A. and {Lanzafame}, A.~C. and {L{\"o}ffler}, W. and {Marchal}, O. and {Marrese}, P.~M. and {Moitinho}, A. and {Muinonen}, K. and {Osborne}, P. and {Pancino}, E. and {Pauwels}, T. and {Recio-Blanco}, A. and {Reyl{\'e}}, C. and {Riello}, M. and {Rimoldini}, L. and {Roegiers}, T. and {Rybizki}, J. and {Sarro}, L.~M. and {Siopis}, C. and {Smith}, M. and {Sozzetti}, A. and {Utrilla}, E. and {van Leeuwen}, M. and {Abbas}, U. and {{\'A}brah{\'a}m}, P. and {Abreu Aramburu}, A. and {Aerts}, C. and {Aguado}, J.~J. and {Ajaj}, M. and {Aldea-Montero}, F. and {Altavilla}, G. and {{\'A}lvarez}, M.~A. and {Alves}, J. and {Anders}, F. and {Anderson}, R.~I. and {Anglada Varela}, E. and {Antoja}, T. and {Baines}, D. and {Baker}, S.~G. and {Balaguer-N{\'u}{\~n}ez}, L. and {Balbinot}, E. and {Balog}, Z. and {Barache}, C. and {Barbato}, D. and {Barros}, M. and {Barstow}, M.~A. and {Bartolom{\'e}}, S. and {Bassilana}, J.-L. and {Bauchet}, N. and {Becciani}, U. and {Bellazzini}, M. and {Berihuete}, A. and {Bernet}, M. and {Bertone}, S. and {Bianchi}, L. and {Binnenfeld}, A. and {Blanco-Cuaresma}, S. and {Blazere}, A. and {Boch}, T. and {Bombrun}, A. and {Bossini}, D. and {Bouquillon}, S. and {Bragaglia}, A. and {Bramante}, L. and {Breedt}, E. and {Bressan}, A. and {Brouillet}, N. and {Brugaletta}, E. and {Bucciarelli}, B. and {Burlacu}, A. and {Butkevich}, A.~G. and {Buzzi}, R. and {Caffau}, E. and {Cancelliere}, R. and {Cantat-Gaudin}, T. and {Carballo}, R. and {Carlucci}, T. and {Carnerero}, M.~I. and {Carrasco}, J.~M. and {Casamiquela}, L. and {Castellani}, M. and {Castro-Ginard}, A. and {Chaoul}, L. and {Charlot}, P. and {Chemin}, L. and {Chiaramida}, V. and {Chiavassa}, A. and {Chornay}, N. and {Comoretto}, G. and {Contursi}, G. and {Cooper}, W.~J. and {Cornez}, T. and {Cowell}, S. and {Crifo}, F. and {Cropper}, M. and {Crosta}, M. and {Crowley}, C. and {Dafonte}, C. and {Dapergolas}, A. and {David}, M. and {David}, P. and {de Laverny}, P. and {De Luise}, F. and {De March}, R.},
        title = "{Gaia Data Release 3. Summary of the content and survey properties}",
      journal = {\aap},
         year = 2023,
        month = jun,
       volume = {674},
          eid = {A1},
        pages = {A1},
          doi = {10.1051/0004-6361/202243940},
archivePrefix = {arXiv},
       eprint = {2208.00211},
 primaryClass = {astro-ph.GA},
       adsurl = {https://ui.adsabs.harvard.edu/abs/2023A&A...674A...1G}
}

@ARTICLE{Frankel2018,
       author = {{Frankel}, Neige and {Rix}, Hans-Walter and {Ting}, Yuan-Sen and {Ness}, Melissa and {Hogg}, David W.},
        title = "{Measuring Radial Orbit Migration in the Galactic Disk}",
      journal = {\apj},
         year = 2018,
        month = oct,
       volume = {865},
       number = {2},
          eid = {96},
        pages = {96},
          doi = {10.3847/1538-4357/aadba5},
archivePrefix = {arXiv},
       eprint = {1805.09198},
 primaryClass = {astro-ph.GA},
       adsurl = {https://ui.adsabs.harvard.edu/abs/2018ApJ...865...96F}
}

@ARTICLE{Bertelli2001,
       author = {{Bertelli}, G. and {Nasi}, E.},
        title = "{Star Formation History in the Solar Vicinity}",
      journal = {\aj},
         year = 2001,
        month = feb,
       volume = {121},
       number = {2},
        pages = {1013-1023},
          doi = {10.1086/318781},
       adsurl = {https://ui.adsabs.harvard.edu/abs/2001AJ....121.1013B}
}

@ARTICLE{Schroeder2003,
       author = {{Schr{\"o}der}, K.-P. and {Pagel}, B.~E.~J.},
        title = "{Galactic archaeology: initial mass function and depletion in the `thin disc'}",
      journal = {\mnras},
         year = 2003,
        month = aug,
       volume = {343},
       number = {4},
        pages = {1231-1240},
          doi = {10.1046/j.1365-8711.2003.06763.x},
archivePrefix = {arXiv},
       eprint = {astro-ph/0303013},
 primaryClass = {astro-ph},
       adsurl = {https://ui.adsabs.harvard.edu/abs/2003MNRAS.343.1231S}
}

@ARTICLE{Hernandez2000,
       author = {{Hernandez}, X. and {Valls-Gabaud}, David and {Gilmore}, Gerard},
        title = "{The recent star formation history of the Hipparcos solar neighbourhood}",
      journal = {\mnras},
         year = 2000,
        month = aug,
       volume = {316},
       number = {3},
        pages = {605-612},
          doi = {10.1046/j.1365-8711.2000.03537.x},
archivePrefix = {arXiv},
       eprint = {astro-ph/0003113},
 primaryClass = {astro-ph},
       adsurl = {https://ui.adsabs.harvard.edu/abs/2000MNRAS.316..605H}
}

@ARTICLE{vergely2002,
       author = {{Vergely}, J. -L. and {K{\"o}ppen}, J. and {Egret}, D. and {Bienaym{\'e}}, O.},
        title = "{An inverse method to interpret colour-magnitude diagrams}",
      journal = {\aap},
         year = 2002,
        month = aug,
       volume = {390},
        pages = {917-929},
          doi = {10.1051/0004-6361:20020334},
archivePrefix = {arXiv},
       eprint = {astro-ph/0203074},
 primaryClass = {astro-ph},
       adsurl = {https://ui.adsabs.harvard.edu/abs/2002A&A...390..917V}
}

@article{cignoni2006,
	title = {Recovering the star formation rate in the solar neighborhood},
	volume = {459},
	issn = {0004-6361, 1432-0746},
	url = {http://www.aanda.org/10.1051/0004-6361:20065645},
	doi = {10.1051/0004-6361:20065645},
	number = {3},
	urldate = {2023-07-24},
	journal = {A\&A},
	author = {Cignoni, M. and Degl'Innocenti, S. and Prada Moroni, P. G. and Shore, S. N.},
	month = dec,
	year = {2006},
	pages = {783--796},
}

@INPROCEEDINGS{bernard2018,
       author = {{Bernard}, Edouard J.},
        title = "{Gaia DR1 completeness within 250 pc \& star formation history of the Solar neighbourhood}",
    booktitle = {Astrometry and Astrophysics in the Gaia Sky},
         year = 2018,
       editor = {{Recio-Blanco}, A. and {de Laverny}, P. and {Brown}, A.~G.~A. and {Prusti}, T.},
       series = {IAU Symposium},
       volume = {330},
        month = apr,
        pages = {148-151},
          doi = {10.1017/S1743921317006159},
archivePrefix = {arXiv},
       eprint = {1801.01427},
 primaryClass = {astro-ph.GA},
       adsurl = {https://ui.adsabs.harvard.edu/abs/2018IAUS..330..148B}
}

@ARTICLE{daltio2021,
       author = {{Dal Tio}, Piero and {Mazzi}, Alessandro and {Girardi}, L{\'e}o and {Barbieri}, Mauro and {Zaggia}, Simone and {Bressan}, Alessandro and {Chen}, Yang and {Costa}, Guglielmo and {Marigo}, Paola},
        title = "{Dissecting the Gaia HR diagram within 200 pc}",
      journal = {\mnras},
         year = 2021,
        month = oct,
       volume = {506},
       number = {4},
        pages = {5681-5697},
          doi = {10.1093/mnras/stab1964},
archivePrefix = {arXiv},
       eprint = {2107.01844},
 primaryClass = {astro-ph.SR},
       adsurl = {https://ui.adsabs.harvard.edu/abs/2021MNRAS.506.5681D}
}

@ARTICLE{alzate2021,
       author = {{Alzate}, Jairo A. and {Bruzual}, Gustavo and {D{\'\i}az-Gonz{\'a}lez}, Daniel J.},
        title = "{Star formation history of the solar neighbourhood as told by Gaia}",
      journal = {\mnras},
         year = 2021,
        month = jan,
       volume = {501},
       number = {1},
        pages = {302-328},
          doi = {10.1093/mnras/staa3576},
archivePrefix = {arXiv},
       eprint = {2011.05732},
 primaryClass = {astro-ph.GA},
       adsurl = {https://ui.adsabs.harvard.edu/abs/2021MNRAS.501..302A}
}

@article{mazzi2023,
       author = {{Mazzi}, Alessandro and {Girardi}, L{\'e}o and {Trabucchi}, Michele and {Dalcanton}, Julianne J. and {Luger}, Rodrigo and {Marigo}, Paola and {Miglio}, Andrea and {Costa}, Guglielmo and {Chen}, Yang and {Pastorelli}, Giada and {Fouesneau}, Morgan and {Zaggia}, Simone and {Bressan}, Alessandro and {Dal Tio}, Piero},
        title = "{Dissecting the Gaia HR diagram - II. The vertical structure of the star formation history across the solar cylinder}",
      journal = {\mnras},
         year = 2024,
        month = jan,
       volume = {527},
       number = {1},
        pages = {583-602},
          doi = {10.1093/mnras/stad2952},
archivePrefix = {arXiv},
       eprint = {2309.13453},
 primaryClass = {astro-ph.GA},
       adsurl = {https://ui.adsabs.harvard.edu/abs/2024MNRAS.527..583M}
}

@ARTICLE{gallart2024,
       author = {{Gallart}, Carme and {Surot}, Francisco and {Cassisi}, Santi and {Fern{\'a}ndez-Alvar}, Emma and {Mirabal}, David and {Rivero}, Alicia and {Ruiz-Lara}, Tom{\'a}s and {Santos-Torres}, Judith and {Aznar-Menargues}, Guillem and {Battaglia}, Giuseppina and {Queiroz}, Anna B. and {Monelli}, Matteo and {Vasiliev}, Eugene and {Chiappini}, Cristina and {Helmi}, Amina and {Hill}, Vanessa and {Massari}, Davide and {Thomas}, Guillaume F.},
        title = "{Chronology of our Galaxy from Gaia colour-magnitude diagram fitting (ChronoGal). I. The formation and evolution of the thin disc from the Gaia Catalogue of Nearby Stars}",
      journal = {\aap},
         year = 2024,
        month = jul,
       volume = {687},
          eid = {A168},
        pages = {A168},
          doi = {10.1051/0004-6361/202349078},
archivePrefix = {arXiv},
       eprint = {2402.09399},
 primaryClass = {astro-ph.GA},
       adsurl = {https://ui.adsabs.harvard.edu/abs/2024A&A...687A.168G}
}

@ARTICLE{ruizlara2020,
       author = {{Ruiz-Lara}, Tom{\'a}s and {Gallart}, Carme and {Bernard}, Edouard J. and {Cassisi}, Santi},
        title = "{The recurrent impact of the Sagittarius dwarf on the star formation history of the Milky Way}",
      journal = {Nature Astronomy},

year = 2020,
        month = may,
       volume = {4},
        pages = {965-973},
          doi = {10.1038/s41550-020-1097-0},
archivePrefix = {arXiv},
       eprint = {2003.12577},
 primaryClass = {astro-ph.GA},
       adsurl = {https://ui.adsabs.harvard.edu/abs/2020NatAs...4..965R}
}

@article{Winget1987,
       author = {{Winget}, D.~E. and {Hansen}, C.~J. and {Liebert}, James and {van Horn}, H.~M. and {Fontaine}, G. and {Nather}, R.~E. and {Kepler}, S.~O. and {Lamb}, D.~Q.},
        title = "{An Independent Method for Determining the Age of the Universe}",
      journal = {\apjl},
         year = 1987,
        month = apr,
       volume = {315},
        pages = {L77},
          doi = {10.1086/184864},
       adsurl = {https://ui.adsabs.harvard.edu/abs/1987ApJ...315L..77W}
}

@article{Liebert1988,
       author = {{Liebert}, James and {Dahn}, Conard C. and {Monet}, David G.},
        title = "{The Luminosity Function of White Dwarfs}",
      journal = {\apj},
         year = 1988,
        month = sep,
       volume = {332},
        pages = {891},
          doi = {10.1086/166699},
       adsurl = {https://ui.adsabs.harvard.edu/abs/1988ApJ...332..891L}
}

@article{Oswalt1996,
  author = {Oswalt, T. D. and Smith, J. A. and Wood, M. A. and Hintzen, P.},
  title  = {A lower limit of 9.5 Gyr on the age of the Galactic disk from the oldest white dwarf stars},
  journal= {Nature},
  year   = {1996}, volume = {382}, pages = {692--694}, doi = {10.1038/382692a0}
}

@article{Leggett1998,
       author = {{Leggett}, S.~K. and {Ruiz}, Maria Teresa and {Bergeron}, P.},
        title = "{The Cool White Dwarf Luminosity Function and the Age of the Galactic Disk}",
      journal = {\apj},
         year = 1998,
        month = apr,
       volume = {497},
       number = {1},
        pages = {294-302},
          doi = {10.1086/305463},
       adsurl = {https://ui.adsabs.harvard.edu/abs/1998ApJ...497..294L}
}

@ARTICLE{Harris2006,
       author = {{Harris}, Hugh C. and {Munn}, Jeffrey A. and {Kilic}, Mukremin and {Liebert}, James and {Williams}, Kurtis A. and {von Hippel}, Ted and {Levine}, Stephen E. and {Monet}, David G. and {Eisenstein}, Daniel J. and {Kleinman}, S.~J. and {Metcalfe}, T.~S. and {Nitta}, Atsuko and {Winget}, D.~E. and {Brinkmann}, J. and {Fukugita}, Masataka and {Knapp}, G.~R. and {Lupton}, Robert H. and {Smith}, J. Allyn and {Schneider}, Donald P.},
        title = "{The White Dwarf Luminosity Function from Sloan Digital Sky Survey Imaging Data}",
      journal = {\aj},
         year = 2006,
        month = jan,
       volume = {131},
       number = {1},
        pages = {571-581},
          doi = {10.1086/497966},
archivePrefix = {arXiv},
       eprint = {astro-ph/0510820},
 primaryClass = {astro-ph},
       adsurl = {https://ui.adsabs.harvard.edu/abs/2006AJ....131..571H}
}

@ARTICLE{Giammichele2012,
       author = {{Giammichele}, N. and {Bergeron}, P. and {Dufour}, P.},
        title = "{Know Your Neighborhood: A Detailed Model Atmosphere Analysis of Nearby White Dwarfs}",
      journal = {\apjs},
         year = 2012,
        month = apr,
       volume = {199},
       number = {2},
          eid = {29},
        pages = {29},
          doi = {10.1088/0067-0049/199/2/29},
archivePrefix = {arXiv},
       eprint = {1202.5581},
 primaryClass = {astro-ph.SR},
       adsurl = {https://ui.adsabs.harvard.edu/abs/2012ApJS..199...29G}
}

@ARTICLE{Rowell2013,
       author = {{Rowell}, N.},
        title = "{The star formation history of the solar neighbourhood from the white dwarf luminosity function}",
      journal = {\mnras},
         year = 2013,
        month = sep,
       volume = {434},
       number = {2},
        pages = {1549-1564},
          doi = {10.1093/mnras/stt1110},
archivePrefix = {arXiv},
       eprint = {1306.4195},
 primaryClass = {astro-ph.GA},
       adsurl = {https://ui.adsabs.harvard.edu/abs/2013MNRAS.434.1549R}
}

@article{Tremblay2014,
       author = {{Tremblay}, P.-E. and {Kalirai}, J.~S. and {Soderblom}, D.~R. and {Cignoni}, M. and {Cummings}, J.},
        title = "{White Dwarf Cosmochronology in the Solar Neighborhood}",
      journal = {\apj},
         year = 2014,
        month = aug,
       volume = {791},
       number = {2},
          eid = {92},
        pages = {92},
          doi = {10.1088/0004-637X/791/2/92},
archivePrefix = {arXiv},
       eprint = {1406.5173},
 primaryClass = {astro-ph.SR},
       adsurl = {https://ui.adsabs.harvard.edu/abs/2014ApJ...791...92T}
}

@article{Lam2019,
       author = {{Lam}, Marco C. and {Hambly}, Nigel C. and {Rowell}, Nicholas and {Chambers}, Kenneth C. and {Goldman}, Bertrand and {Hodapp}, Klaus W. and {Kaiser}, Nick and {Kudritzki}, Rolf-Peter and {Magnier}, Eugene A. and {Tonry}, John L. and {Wainscoat}, Richard J. and {Waters}, Christopher},
        title = "{The white dwarf luminosity functions from the Pan-STARRS 1 3{\ensuremath{\pi}} Steradian Survey}",
      journal = {\mnras},
         year = 2019,
        month = jan,
       volume = {482},
       number = {1},
        pages = {715-731},
          doi = {10.1093/mnras/sty2710},
archivePrefix = {arXiv},
       eprint = {1810.01798},
 primaryClass = {astro-ph.GA},
       adsurl = {https://ui.adsabs.harvard.edu/abs/2019MNRAS.482..715L}
}

@article{JimenezEsteban2018,
       author = {{Jim{\'e}nez-Esteban}, F.~M. and {Torres}, S. and {Rebassa-Mansergas}, A. and {Skorobogatov}, G. and {Solano}, E. and {Cantero}, C. and {Rodrigo}, C.},
        title = "{A white dwarf catalogue from Gaia-DR2 and the Virtual Observatory}",
      journal = {\mnras},
         year = 2018,
        month = nov,
       volume = {480},
       number = {4},
        pages = {4505-4518},
          doi = {10.1093/mnras/sty2120},
archivePrefix = {arXiv},
       eprint = {1807.02559},
 primaryClass = {astro-ph.SR},
       adsurl = {https://ui.adsabs.harvard.edu/abs/2018MNRAS.480.4505J}
}

@ARTICLE{GentileFusillo2019,
       author = {{Gentile Fusillo}, Nicola Pietro and {Tremblay}, Pier-Emmanuel and {G{\"a}nsicke}, Boris T. and {Manser}, Christopher J. and {Cunningham}, Tim and {Cukanovaite}, Elena and {Hollands}, Mark and {Marsh}, Thomas and {Raddi}, Roberto and {Jordan}, Stefan and {Toonen}, Silvia and {Geier}, Stephan and {Barstow}, Martin and {Cummings}, Jeffrey D.},
        title = "{A Gaia Data Release 2 catalogue of white dwarfs and a comparison with SDSS}",
      journal = {\mnras},
         year = 2019,
        month = feb,
       volume = {482},
       number = {4},
        pages = {4570-4591},
          doi = {10.1093/mnras/sty3016},
archivePrefix = {arXiv},
       eprint = {1807.03315},
 primaryClass = {astro-ph.SR},
       adsurl = {https://ui.adsabs.harvard.edu/abs/2019MNRAS.482.4570G}
}

@ARTICLE{Cukanovaite2023,
       author = {{Cukanovaite}, E. and {Tremblay}, P.-E. and {Toonen}, S. and {Temmink}, K.~D. and {Manser}, Christopher J. and {O'Brien}, M.~W. and {McCleery}, J.},
        title = "{Local stellar formation history from the 40 pc white dwarf sample}",
      journal = {\mnras},
         year = 2023,
        month = jun,
       volume = {522},
       number = {2},
        pages = {1643-1661},
          doi = {10.1093/mnras/stad1020},
archivePrefix = {arXiv},
       eprint = {2209.13919},
 primaryClass = {astro-ph.SR},
       adsurl = {https://ui.adsabs.harvard.edu/abs/2023MNRAS.522.1643C}
}

@ARTICLE{Roberts2025,
       author = {{Roberts}, Emily K. and {Tremblay}, Pier-Emmanuel and {O'Brien}, Mairi W. and {B{\'e}dard}, Antoine and {Cunningham}, Tim and {Byrne}, Conor M. and {Cukanovaite}, Elena},
        title = "{Comparison of methods used to derive the Galactic star formation history from white dwarf samples}",
      journal = {\mnras},
         year = 2025,
        month = apr,
       volume = {538},
       number = {4},
        pages = {2548-2561},
          doi = {10.1093/mnras/staf434},
archivePrefix = {arXiv},
       eprint = {2502.09579},
 primaryClass = {astro-ph.SR},
       adsurl = {https://ui.adsabs.harvard.edu/abs/2025MNRAS.538.2548R}
}

@ARTICLE{Obrien2024,
       author = {{O'Brien}, Mairi W. and {Tremblay}, P.-E. and {Klein}, B.~L. and {Koester}, D. and {Melis}, C. and {B{\'e}dard}, A. and {Cukanovaite}, E. and {Cunningham}, T. and {Doyle}, A.~E. and {G{\"a}nsicke}, B.~T. and {Gentile Fusillo}, N.~P. and {Hollands}, M.~A. and {McCleery}, J. and {Pelisoli}, I. and {Toonen}, S. and {Weinberger}, A.~J. and {Zuckerman}, B.},
        title = "{The 40 pc sample of white dwarfs from Gaia}",
      journal = {\mnras},
         year = 2024,
        month = jan,
       volume = {527},
       number = {3},
        pages = {8687-8705},
          doi = {10.1093/mnras/stad3773},
archivePrefix = {arXiv},
       eprint = {2312.02735},
 primaryClass = {astro-ph.SR},
       adsurl = {https://ui.adsabs.harvard.edu/abs/2024MNRAS.527.8687O}
}

@ARTICLE{LamRowell2025,
       author = {{Lam}, M.~C. and {Rowell}, N. and {Yeung}, H.~W.},
        title = "{A new method to retrieve the star formation history from white dwarf luminosity functions ─ an application to the Gaia catalogue of nearby stars}",
      journal = {\mnras},
         year = 2026,
        month = feb,
       volume = {546},
       number = {1},
          eid = {staf2266},
        pages = {staf2266},
          doi = {10.1093/mnras/staf2266},
archivePrefix = {arXiv},
       eprint = {2503.14394},
 primaryClass = {astro-ph.SR},
       adsurl = {https://ui.adsabs.harvard.edu/abs/2026MNRAS.546f2266L}
}

@ARTICLE{cignoni2015,
       author = {{Cignoni}, M. and {Sabbi}, E. and {van der Marel}, R.~P. and {Tosi}, M. and {Zaritsky}, D. and {Anderson}, J. and {Lennon}, D.~J. and {Aloisi}, A. and {de Marchi}, G. and {Gouliermis}, D.~A. and {Grebel}, E.~K. and {Smith}, L.~J. and {Zeidler}, P.},
        title = "{Hubble Tarantula Treasury Project. II. The Star-formation History of the Starburst Region NGC 2070 in 30 Doradus}",
      journal = {\apj},
         year = 2015,
        month = oct,
       volume = {811},
       number = {2},
          eid = {76},
        pages = {76},
          doi = {10.1088/0004-637X/811/2/76},
archivePrefix = {arXiv},
       eprint = {1505.04799},
 primaryClass = {astro-ph.SR},
       adsurl = {https://ui.adsabs.harvard.edu/abs/2015ApJ...811...76C}
}

@ARTICLE{Sacchi2018,
       author = {{Sacchi}, E. and {Cignoni}, M. and {Aloisi}, A. and {Tosi}, M. and {Calzetti}, D. and {Lee}, J.~C. and {Adamo}, A. and {Annibali}, F. and {Dale}, D.~A. and {Elmegreen}, B.~G. and {Gouliermis}, D.~A. and {Grasha}, K. and {Grebel}, E.~K. and {Hunter}, D.~A. and {Sabbi}, E. and {Smith}, L.~J. and {Thilker}, D.~A. and {Ubeda}, L. and {Whitmore}, B.~C.},
        title = "{Star Formation Histories of the LEGUS Dwarf Galaxies. II. Spatially Resolved Star Formation History of the Magellanic Irregular NGC 4449}",
      journal = {\apj},
         year = 2018,
        month = apr,
       volume = {857},
       number = {1},
          eid = {63},
        pages = {63},
          doi = {10.3847/1538-4357/aab844},
archivePrefix = {arXiv},
       eprint = {1803.08041},
 primaryClass = {astro-ph.GA},
       adsurl = {https://ui.adsabs.harvard.edu/abs/2018ApJ...857...63S}
}

@ARTICLE{Lindegren2021,
       author = {{Lindegren}, L. and {Bastian}, U. and {Biermann}, M. and {Bombrun}, A. and {de Torres}, A. and {Gerlach}, E. and {Geyer}, R. and {Hern{\'a}ndez}, J. and {Hilger}, T. and {Hobbs}, D. and {Klioner}, S.~A. and {Lammers}, U. and {McMillan}, P.~J. and {Ramos-Lerate}, M. and {Steidelm{\"u}ller}, H. and {Stephenson}, C.~A. and {van Leeuwen}, F.},
        title = "{Gaia Early Data Release 3. Parallax bias versus magnitude, colour, and position}",
      journal = {\aap},
         year = 2021,
        month = may,
       volume = {649},
          eid = {A4},
        pages = {A4},
          doi = {10.1051/0004-6361/202039653},
archivePrefix = {arXiv},
       eprint = {2012.01742},
 primaryClass = {astro-ph.IM},
       adsurl = {https://ui.adsabs.harvard.edu/abs/2021A&A...649A...4L}
}

@ARTICLE{riello2021,
       author = {{Riello}, M. and {De Angeli}, F. and {Evans}, D.~W. and {Montegriffo}, P. and {Carrasco}, J.~M. and {Busso}, G. and {Palaversa}, L. and {Burgess}, P.~W. and {Diener}, C. and {Davidson}, M. and {Rowell}, N. and {Fabricius}, C. and {Jordi}, C. and {Bellazzini}, M. and {Pancino}, E. and {Harrison}, D.~L. and {Cacciari}, C. and {van Leeuwen}, F. and {Hambly}, N.~C. and {Hodgkin}, S.~T. and {Osborne}, P.~J. and {Altavilla}, G. and {Barstow}, M.~A. and {Brown}, A.~G.~A. and {Castellani}, M. and {Cowell}, S. and {De Luise}, F. and {Gilmore}, G. and {Giuffrida}, G. and {Hidalgo}, S. and {Holland}, G. and {Marinoni}, S. and {Pagani}, C. and {Piersimoni}, A.~M. and {Pulone}, L. and {Ragaini}, S. and {Rainer}, M. and {Richards}, P.~J. and {Sanna}, N. and {Walton}, N.~A. and {Weiler}, M. and {Yoldas}, A.},
        title = "{Gaia Early Data Release 3. Photometric content and validation}",
      journal = {\aap},
         year = 2021,
        month = may,
       volume = {649},
          eid = {A3},
        pages = {A3},
          doi = {10.1051/0004-6361/202039587},
archivePrefix = {arXiv},
       eprint = {2012.01916},
 primaryClass = {astro-ph.IM},
       adsurl = {https://ui.adsabs.harvard.edu/abs/2021A&A...649A...3R}
}

@ARTICLE{vergely2022,
       author = {{Vergely}, J.~L. and {Lallement}, R. and {Cox}, N.~L.~J.},
        title = "{Three-dimensional extinction maps: Inverting inter-calibrated extinction catalogues}",
      journal = {\aap},
         year = 2022,
        month = aug,
       volume = {664},
          eid = {A174},
        pages = {A174},
          doi = {10.1051/0004-6361/202243319},
archivePrefix = {arXiv},
       eprint = {2205.09087},
 primaryClass = {astro-ph.GA},
       adsurl = {https://ui.adsabs.harvard.edu/abs/2022A&A...664A.174V}
}

@ARTICLE{tosi1990,
       author = {{Tosi}, M. and {Greggio}, L. and {Marconi}, G. and {Focardi}, P.},
        title = "{Star Formation in Dwarf Irregular Galaxies: Sextans B}",
      journal = {\aj},
         year = 1991,
        month = sep,
       volume = {102},
        pages = {951},
          doi = {10.1086/115925},
       adsurl = {https://ui.adsabs.harvard.edu/abs/1991AJ....102..951T}
}

@ARTICLE{tolstoy1996,
       author = {{Tolstoy}, Eline and {Saha}, Abhijit},
        title = "{The Interpretation of Color-Magnitude Diagrams through Numerical Simulation and Bayesian Inference}",
      journal = {\apj},
         year = 1996,
        month = may,
       volume = {462},
        pages = {672},
          doi = {10.1086/177181},
       adsurl = {https://ui.adsabs.harvard.edu/abs/1996ApJ...462..672T}
}

@ARTICLE{dolphin1997,
       author = {{Dolphin}, Andrew},
        title = "{A new method to determine star formation histories of nearby galaxies}",
      journal = {\na},
         year = 1997,
        month = nov,
       volume = {2},
       number = {5},
        pages = {397-409},
          doi = {10.1016/S1384-1076(97)00029-8},
       adsurl = {https://ui.adsabs.harvard.edu/abs/1997NewA....2..397D}
}

@ARTICLE{dolphin2002,
       author = {{Dolphin}, A.~E.},
        title = "{Numerical methods of star formation history measurement and applications to seven dwarf spheroidals}",
      journal = {\mnras},
         year = 2002,
        month = may,
       volume = {332},
       number = {1},
        pages = {91-108},
          doi = {10.1046/j.1365-8711.2002.05271.x},
archivePrefix = {arXiv},
       eprint = {astro-ph/0112331},
 primaryClass = {astro-ph},
       adsurl = {https://ui.adsabs.harvard.edu/abs/2002MNRAS.332...91D}
}

@ARTICLE{cignoni&tosi2010,
       author = {{Cignoni}, Michele and {Tosi}, Monica},
        title = "{Star Formation Histories of Dwarf Galaxies from the Colour-Magnitude Diagrams of Their Resolved Stellar Populations}",
      journal = {Advances in Astronomy},
         year = 2010,
        month = jan,
       volume = {2010},
          eid = {158568},
        pages = {158568},
          doi = {10.1155/2010/158568},
archivePrefix = {arXiv},
       eprint = {0909.4234},
 primaryClass = {astro-ph.GA},
       adsurl = {https://ui.adsabs.harvard.edu/abs/2010AdAst2010E...3C}
}

@ARTICLE{aparicio&hidalgo2009,
       author = {{Aparicio}, Antonio and {Hidalgo}, Sebastian L.},
        title = "{IAC-pop: Finding the Star Formation History of Resolved Galaxies}",
      journal = {\aj},
         year = 2009,
        month = aug,
       volume = {138},
       number = {2},
        pages = {558-567},
          doi = {10.1088/0004-6256/138/2/558},
archivePrefix = {arXiv},
       eprint = {0906.0712},
 primaryClass = {astro-ph.CO},
       adsurl = {https://ui.adsabs.harvard.edu/abs/2009AJ....138..558A}
}

@ARTICLE{weisz2012,
       author = {{Weisz}, Daniel R. and {Zucker}, Daniel B. and {Dolphin}, Andrew E. and {Martin}, Nicolas F. and {de Jong}, Jelte T.~A. and {Holtzman}, Jon A. and {Dalcanton}, Julianne J. and {Gilbert}, Karoline M. and {Williams}, Benjamin F. and {Bell}, Eric F. and {Belokurov}, Vasily and {Evans}, N. Wyn},
        title = "{The Star Formation History of Leo T from Hubble Space Telescope Imaging}",
      journal = {\apj},
         year = 2012,
        month = apr,
       volume = {748},
       number = {2},
          eid = {88},
        pages = {88},
          doi = {10.1088/0004-637X/748/2/88},
archivePrefix = {arXiv},
       eprint = {1201.4859},
 primaryClass = {astro-ph.CO},
       adsurl = {https://ui.adsabs.harvard.edu/abs/2012ApJ...748...88W}
}

@ARTICLE{ruizlara2018,
       author = {{Ruiz-Lara}, T. and {Gallart}, C. and {Beasley}, M. and {Monelli}, M. and {Bernard}, E.~J. and {Battaglia}, G. and {S{\'a}nchez-Bl{\'a}zquez}, P. and {Florido}, E. and {P{\'e}rez}, I. and {Mart{\'\i}n-Navarro}, I.},
        title = "{Integrated-light analyses vs. colour-magnitude diagrams. II. Leo A: an extremely young dwarf in the Local Group}",
      journal = {\aap},
         year = 2018,
        month = sep,
       volume = {617},
          eid = {A18},
        pages = {A18},
          doi = {10.1051/0004-6361/201732398},
archivePrefix = {arXiv},
       eprint = {1805.04323},
 primaryClass = {astro-ph.GA},
       adsurl = {https://ui.adsabs.harvard.edu/abs/2018A&A...617A..18R}
}

@article{mor2019,
	title = {Gaia {DR2} reveals a star formation burst in the disc 2-3 {Gyr} ago},
	volume = {624},
	issn = {0004-6361, 1432-0746},
	doi = {10.1051/0004-6361/201935105},
	language = {en},
	urldate = {2023-05-04},
	journal = {A\&A},
	author = {Mor, R. and Robin, A. C. and Figueras, F. and Roca-Fàbrega, S. and Luri, X.},
	month = apr,
	year = {2019},
	pages = {L1},
}

@ARTICLE{kroupa2001,
       author = {{Kroupa}, Pavel},
        title = "{On the variation of the initial mass function}",
      journal = {\mnras},
         year = 2001,
        month = apr,
       volume = {322},
       number = {2},
        pages = {231-246},
          doi = {10.1046/j.1365-8711.2001.04022.x},
archivePrefix = {arXiv},
       eprint = {astro-ph/0009005},
 primaryClass = {astro-ph},
       adsurl = {https://ui.adsabs.harvard.edu/abs/2001MNRAS.322..231K}
}

@ARTICLE{Bressan2012,
       author = {{Bressan}, Alessandro and {Marigo}, Paola and {Girardi}, L{\'e}o. and {Salasnich}, Bernardo and {Dal Cero}, Claudia and {Rubele}, Stefano and {Nanni}, Ambra},
        title = "{PARSEC: stellar tracks and isochrones with the PAdova and TRieste Stellar Evolution Code}",
      journal = {\mnras},
         year = 2012,
        month = nov,
       volume = {427},
       number = {1},
        pages = {127-145},
          doi = {10.1111/j.1365-2966.2012.21948.x},
archivePrefix = {arXiv},
       eprint = {1208.4498},
 primaryClass = {astro-ph.SR},
       adsurl = {https://ui.adsabs.harvard.edu/abs/2012MNRAS.427..127B}
}

@ARTICLE{Tang2014,
       author = {{Tang}, Jing and {Bressan}, Alessandro and {Rosenfield}, Philip and {Slemer}, Alessandra and {Marigo}, Paola and {Girardi}, L{\'e}o and {Bianchi}, Luciana},
        title = "{New PARSEC evolutionary tracks of massive stars at low metallicity: testing canonical stellar evolution in nearby star-forming dwarf galaxies}",
      journal = {\mnras},
         year = 2014,
        month = dec,
       volume = {445},
       number = {4},
        pages = {4287-4305},
          doi = {10.1093/mnras/stu2029},
archivePrefix = {arXiv},
       eprint = {1410.1745},
 primaryClass = {astro-ph.SR},
       adsurl = {https://ui.adsabs.harvard.edu/abs/2014MNRAS.445.4287T}
}

@ARTICLE{Marigo2017,
       author = {{Marigo}, Paola and {Girardi}, L{\'e}o and {Bressan}, Alessandro and {Rosenfield}, Philip and {Aringer}, Bernhard and {Chen}, Yang and {Dussin}, Marco and {Nanni}, Ambra and {Pastorelli}, Giada and {Rodrigues}, Tha{\'\i}se S. and {Trabucchi}, Michele and {Bladh}, Sara and {Dalcanton}, Julianne and {Groenewegen}, Martin A.~T. and {Montalb{\'a}n}, Josefina and {Wood}, Peter R.},
        title = "{A New Generation of PARSEC-COLIBRI Stellar Isochrones Including the TP-AGB Phase}",
      journal = {\apj},
         year = 2017,
        month = jan,
       volume = {835},
       number = {1},
          eid = {77},
        pages = {77},
          doi = {10.3847/1538-4357/835/1/77},
archivePrefix = {arXiv},
       eprint = {1701.08510},
 primaryClass = {astro-ph.SR},
       adsurl = {https://ui.adsabs.harvard.edu/abs/2017ApJ...835...77M}
}

@article{tognelli_cumulative_2015,
       author = {{Tognelli}, E. and {Prada Moroni}, P.~G. and {Degl'Innocenti}, S.},
        title = "{Cumulative theoretical uncertainties in lithium depletion boundary age}",
      journal = {\mnras},
         year = 2015,
        month = jun,
       volume = {449},
       number = {4},
        pages = {3741-3754},
          doi = {10.1093/mnras/stv577},
       adsurl = {https://ui.adsabs.harvard.edu/abs/2015MNRAS.449.3741T}
}

@ARTICLE{cignoni2009,
       author = {{Cignoni}, M. and {Sabbi}, E. and {Nota}, A. and {Tosi}, M. and {Degl'Innocenti}, S. and {Moroni}, P.~G. Prada and {Angeretti}, L. and {Carlson}, Lynn Redding and {Gallagher}, J. and {Meixner}, M. and {Sirianni}, M. and {Smith}, L.~J.},
        title = "{Star Formation History in the Small Magellanic Cloud: The Case of NGC 602}",
      journal = {\aj},
         year = 2009,
        month = mar,
       volume = {137},
       number = {3},
        pages = {3668-3684},
          doi = {10.1088/0004-6256/137/3/3668},
archivePrefix = {arXiv},
       eprint = {0901.1237},
 primaryClass = {astro-ph.GA},
       adsurl = {https://ui.adsabs.harvard.edu/abs/2009AJ....137.3668C}
}

@ARTICLE{cash79,
       author = {{Cash}, W.},
        title = "{Parameter estimation in astronomy through application of the likelihood ratio.}",
      journal = {\apj},
         year = 1979,
        month = mar,
       volume = {228},
        pages = {939-947},
          doi = {10.1086/156922},
       adsurl = {https://ui.adsabs.harvard.edu/abs/1979ApJ...228..939C}
}

@ARTICLE{apogee2022,
       author = {{Abdurro'uf} and {Accetta}, Katherine and {Aerts}, Conny and {Silva Aguirre}, V{\'\i}ctor and {Ahumada}, Romina and {Ajgaonkar}, Nikhil and {Filiz Ak}, N. and {Alam}, Shadab and {Allende Prieto}, Carlos and {Almeida}, Andr{\'e}s and {Anders}, Friedrich and {Anderson}, Scott F. and {Andrews}, Brett H. and {Anguiano}, Borja and {Aquino-Ort{\'\i}z}, Erik and {Arag{\'o}n-Salamanca}, Alfonso and {Argudo-Fern{\'a}ndez}, Maria and {Ata}, Metin and {Aubert}, Marie and {Avila-Reese}, Vladimir and {Badenes}, Carles and {Barb{\'a}}, Rodolfo H. and {Barger}, Kat and {Barrera-Ballesteros}, Jorge K. and {Beaton}, Rachael L. and {Beers}, Timothy C. and {Belfiore}, Francesco and {Bender}, Chad F. and {Bernardi}, Mariangela and {Bershady}, Matthew A. and {Beutler}, Florian and {Bidin}, Christian Moni and {Bird}, Jonathan C. and {Bizyaev}, Dmitry and {Blanc}, Guillermo A. and {Blanton}, Michael R. and {Boardman}, Nicholas Fraser and {Bolton}, Adam S. and {Boquien}, M{\'e}d{\'e}ric and {Borissova}, Jura and {Bovy}, Jo and {Brandt}, W.~N. and {Brown}, Jordan and {Brownstein}, Joel R. and {Brusa}, Marcella and {Buchner}, Johannes and {Bundy}, Kevin and {Burchett}, Joseph N. and {Bureau}, Martin and {Burgasser}, Adam and {Cabang}, Tuesday K. and {Campbell}, Stephanie and {Cappellari}, Michele and {Carlberg}, Joleen K. and {Wanderley}, F{\'a}bio Carneiro and {Carrera}, Ricardo and {Cash}, Jennifer and {Chen}, Yan-Ping and {Chen}, Wei-Huai and {Cherinka}, Brian and {Chiappini}, Cristina and {Choi}, Peter Doohyun and {Chojnowski}, S. Drew and {Chung}, Haeun and {Clerc}, Nicolas and {Cohen}, Roger E. and {Comerford}, Julia M. and {Comparat}, Johan and {da Costa}, Luiz and {Covey}, Kevin and {Crane}, Jeffrey D. and {Cruz-Gonzalez}, Irene and {Culhane}, Connor and {Cunha}, Katia and {Dai}, Y. Sophia and {Damke}, Guillermo and {Darling}, Jeremy and {Davidson}, Jr., James W. and {Davies}, Roger and {Dawson}, Kyle and {De Lee}, Nathan and {Diamond-Stanic}, Aleksandar M. and {Cano-D{\'\i}az}, Mariana and {S{\'a}nchez}, Helena Dom{\'\i}nguez and {Donor}, John and {Duckworth}, Chris and {Dwelly}, Tom and {Eisenstein}, Daniel J. and {Elsworth}, Yvonne P. and {Emsellem}, Eric and {Eracleous}, Mike and {Escoffier}, Stephanie and {Fan}, Xiaohui and {Farr}, Emily and {Feng}, Shuai and {Fern{\'a}ndez-Trincado}, Jos{\'e} G. and {Feuillet}, Diane and {Filipp}, Andreas and {Fillingham}, Sean P. and {Frinchaboy}, Peter M. and {Fromenteau}, Sebastien and {Galbany}, Llu{\'\i}s and {Garc{\'\i}a}, Rafael A. and {Garc{\'\i}a-Hern{\'a}ndez}, D.~A. and {Ge}, Junqiang and {Geisler}, Doug and {Gelfand}, Joseph and {G{\'e}ron}, Tobias and {Gibson}, Benjamin J. and {Goddy}, Julian and {Godoy-Rivera}, Diego and {Grabowski}, Kathleen and {Green}, Paul J. and {Greener}, Michael and {Grier}, Catherine J. and {Griffith}, Emily and {Guo}, Hong and {Guy}, Julien and {Hadjara}, Massinissa and {Harding}, Paul and {Hasselquist}, Sten and {Hayes}, Christian R. and {Hearty}, Fred and {Hern{\'a}ndez}, Jes{\'u}s and {Hill}, Lewis and {Hogg}, David W. and {Holtzman}, Jon A. and {Horta}, Danny and {Hsieh}, Bau-Ching and {Hsu}, Chin-Hao and {Hsu}, Yun-Hsin and {Huber}, Daniel and {Huertas-Company}, Marc and {Hutchinson}, Brian and {Hwang}, Ho Seong and {Ibarra-Medel}, H{\'e}ctor J. and {Chitham}, Jacob Ider and {Ilha}, Gabriele S. and {Imig}, Julie and {Jaekle}, Will and {Jayasinghe}, Tharindu and {Ji}, Xihan and {Johnson}, Jennifer A. and {Jones}, Amy and {J{\"o}nsson}, Henrik and {Katkov}, Ivan and {Khalatyan}, Dr., Arman and {Kinemuchi}, Karen and {Kisku}, Shobhit and {Knapen}, Johan H. and {Kneib}, Jean-Paul and {Kollmeier}, Juna A. and {Kong}, Miranda and {Kounkel}, Marina and {Kreckel}, Kathryn and {Krishnarao}, Dhanesh and {Lacerna}, Ivan and {Lane}, Richard R. and {Langgin}, Rachel and {Lavender}, Ramon and {Law}, David R. and {Lazarz}, Daniel and {Leung}, Henry W. and {Leung}, Ho-Hin and {Lewis}, Hannah M. and {Li}, Cheng and {Li}, Ran and {Lian}, Jianhui and {Liang}, Fu-Heng and {Lin}, Lihwai and {Lin}, Yen-Ting and {Lin}, Sicheng and {Lintott}, Chris and {Long}, Dan and {Longa-Pe{\~n}a}, Pen{\'e}lope and {L{\'o}pez-Cob{\'a}}, Carlos and {Lu}, Shengdong and {Lundgren}, Britt F. and {Luo}, Yuanze and {Mackereth}, J. Ted and {de la Macorra}, Axel and {Mahadevan}, Suvrath and {Majewski}, Steven R. and {Manchado}, Arturo and {Mandeville}, Travis and {Maraston}, Claudia and {Margalef-Bentabol}, Berta and {Masseron}, Thomas and {Masters}, Karen L. and {Mathur}, Savita and {McDermid}, Richard M. and {Mckay}, Myles and {Merloni}, Andrea and {Merrifield}, Michael and {Meszaros}, Szabolcs and {Miglio}, Andrea and {Di Mille}, Francesco and {Minniti}, Dante and {Minsley}, Rebecca and {Monachesi}, Antonela},
        title = "{The Seventeenth Data Release of the Sloan Digital Sky Surveys: Complete Release of MaNGA, MaStar, and APOGEE-2 Data}",
      journal = {\apjs},
         year = 2022,
        month = apr,
       volume = {259},
       number = {2},
          eid = {35},
        pages = {35},
          doi = {10.3847/1538-4365/ac4414},
archivePrefix = {arXiv},
       eprint = {2112.02026},
 primaryClass = {astro-ph.GA},
       adsurl = {https://ui.adsabs.harvard.edu/abs/2022ApJS..259...35A}
}

@article{Plevne2020,
       author = {{Plevne}, Olcay and {{\"O}nal Ta{\textcommabelow s}}, {\"O}zgecan and {Bilir}, Sel{\c{c}}uk and {Seabroke}, George M.},
        title = "{Multiwavelength Absolute Magnitudes and Colors of Red Clump Stars in the Gaia Era}",
      journal = {\apj},
         year = 2020,
        month = apr,
       volume = {893},
       number = {2},
          eid = {108},
        pages = {108},
          doi = {10.3847/1538-4357/ab80bb},
archivePrefix = {arXiv},
       eprint = {2003.07887},
 primaryClass = {astro-ph.GA},
       adsurl = {https://ui.adsabs.harvard.edu/abs/2020ApJ...893..108P}
}

@article{Reyes2024,
       author = {{Reyes}, Claudia and {Stello}, Dennis and {Hon}, Marc and {Trampedach}, Regner and {Sandquist}, Eric and {Pinsonneault}, Marc H.},
        title = "{Isochrone fitting of the open cluster M67 in the era of Gaia and improved model physics}",
      journal = {\mnras},
         year = 2024,
        month = aug,
       volume = {532},
       number = {2},
        pages = {2860-2874},
          doi = {10.1093/mnras/stae1650},
archivePrefix = {arXiv},
       eprint = {2407.03526},
 primaryClass = {astro-ph.SR},
       adsurl = {https://ui.adsabs.harvard.edu/abs/2024MNRAS.532.2860R}
}

@ARTICLE{Chriss2025,
       author = {{Chriss}, Abigail R. and {Worthey}, Guy},
        title = "{On the Age Calibration of Open Clusters Using Red Clump Stars}",
      journal = {\aj},
         year = 2025,
        month = feb,
       volume = {169},
       number = {2},
          eid = {81},
        pages = {81},
          doi = {10.3847/1538-3881/ada1c4},
archivePrefix = {arXiv},
       eprint = {2402.18538},
 primaryClass = {astro-ph.GA},
       adsurl = {https://ui.adsabs.harvard.edu/abs/2025AJ....169...81C}
}

@ARTICLE{gentilefusillo2021,
       author = {{Gentile Fusillo}, N.~P. and {Tremblay}, P. -E. and {Cukanovaite}, E. and {Vorontseva}, A. and {Lallement}, R. and {Hollands}, M. and {G{\"a}nsicke}, B.~T. and {Burdge}, K.~B. and {McCleery}, J. and {Jordan}, S.},
        title = "{A catalogue of white dwarfs in Gaia EDR3}",
      journal = {\mnras},
         year = 2021,
        month = dec,
       volume = {508},
       number = {3},
        pages = {3877-3896},
          doi = {10.1093/mnras/stab2672},
archivePrefix = {arXiv},
       eprint = {2106.07669},
 primaryClass = {astro-ph.SR},
       adsurl = {https://ui.adsabs.harvard.edu/abs/2021MNRAS.508.3877G}
}

@ARTICLE{fabricius2021,
       author = {{Fabricius}, C. and {Luri}, X. and {Arenou}, F. and {Babusiaux}, C. and {Helmi}, A. and {Muraveva}, T. and {Reyl{\'e}}, C. and {Spoto}, F. and {Vallenari}, A. and {Antoja}, T. and {Balbinot}, E. and {Barache}, C. and {Bauchet}, N. and {Bragaglia}, A. and {Busonero}, D. and {Cantat-Gaudin}, T. and {Carrasco}, J.~M. and {Diakit{\'e}}, S. and {Fabrizio}, M. and {Figueras}, F. and {Garcia-Gutierrez}, A. and {Garofalo}, A. and {Jordi}, C. and {Kervella}, P. and {Khanna}, S. and {Leclerc}, N. and {Licata}, E. and {Lambert}, S. and {Marrese}, P.~M. and {Masip}, A. and {Ramos}, P. and {Robichon}, N. and {Robin}, A.~C. and {Romero-G{\'o}mez}, M. and {Rubele}, S. and {Weiler}, M.},
        title = "{Gaia Early Data Release 3. Catalogue validation}",
      journal = {\aap},
         year = 2021,
        month = may,
       volume = {649},
          eid = {A5},
        pages = {A5},
          doi = {10.1051/0004-6361/202039834},
archivePrefix = {arXiv},
       eprint = {2012.06242},
 primaryClass = {astro-ph.GA},
       adsurl = {https://ui.adsabs.harvard.edu/abs/2021A&A...649A...5F}
}

@ARTICLE{Cunningham2024,
       author = {{Cunningham}, Tim and {Tremblay}, Pier-Emmanuel and {W. O'Brien}, Mairi},
        title = "{Initial-final mass relation from white dwarfs within 40 pc}",
      journal = {\mnras},
         year = 2024,
        month = jan,
       volume = {527},
       number = {2},
        pages = {3602-3611},
          doi = {10.1093/mnras/stad3275},
archivePrefix = {arXiv},
       eprint = {2310.15410},
 primaryClass = {astro-ph.SR},
       adsurl = {https://ui.adsabs.harvard.edu/abs/2024MNRAS.527.3602C}
}

@ARTICLE{esteban2023,
       author = {{Jim{\'e}nez-Esteban}, F.~M. and {Torres}, S. and {Rebassa-Mansergas}, A. and {Cruz}, P. and {Murillo-Ojeda}, R. and {Solano}, E. and {Rodrigo}, C. and {Camisassa}, M.~E.},
        title = "{Spectral classification of the 100 pc white dwarf population from Gaia-DR3 and the virtual observatory}",
      journal = {\mnras},
         year = 2023,
        month = feb,
       volume = {518},
       number = {4},
        pages = {5106-5122},
          doi = {10.1093/mnras/stac3382},
archivePrefix = {arXiv},
       eprint = {2211.08852},
 primaryClass = {astro-ph.SR},
       adsurl = {https://ui.adsabs.harvard.edu/abs/2023MNRAS.518.5106J}
}

@ARTICLE{Blouin2020,
       author = {{Blouin}, Simon and {Shaffer}, Nathaniel R. and {Saumon}, Didier and {Starrett}, Charles E.},
        title = "{New Conductive Opacities for White Dwarf Envelopes}",
      journal = {\apj},
         year = 2020,
        month = aug,
       volume = {899},
       number = {1},
          eid = {46},
        pages = {46},
          doi = {10.3847/1538-4357/ab9e75},
archivePrefix = {arXiv},
       eprint = {2006.16390},
 primaryClass = {astro-ph.SR},
       adsurl = {https://ui.adsabs.harvard.edu/abs/2020ApJ...899...46B}
}

@ARTICLE{Cassisi2007,
       author = {{Cassisi}, S. and {Potekhin}, A.~Y. and {Pietrinferni}, A. and {Catelan}, M. and {Salaris}, M.},
        title = "{Updated Electron-Conduction Opacities: The Impact on Low-Mass Stellar Models}",
      journal = {\apj},
         year = 2007,
        month = jun,
       volume = {661},
       number = {2},
        pages = {1094-1104},
          doi = {10.1086/516819},
archivePrefix = {arXiv},
       eprint = {astro-ph/0703011},
 primaryClass = {astro-ph},
       adsurl = {https://ui.adsabs.harvard.edu/abs/2007ApJ...661.1094C}
}

@ARTICLE{salaris2022,
       author = {{Salaris}, Maurizio and {Cassisi}, Santi and {Pietrinferni}, Adriano and {Hidalgo}, Sebastian},
        title = "{The updated BASTI stellar evolution models and isochrones - III. White dwarfs}",
      journal = {\mnras},
         year = 2022,
        month = feb,
       volume = {509},
       number = {4},
        pages = {5197-5208},
          doi = {10.1093/mnras/stab3359},
archivePrefix = {arXiv},
       eprint = {2111.09285},
 primaryClass = {astro-ph.SR},
       adsurl = {https://ui.adsabs.harvard.edu/abs/2022MNRAS.509.5197S}
}

@ARTICLE{bauer2026,
       author = {{Bauer}, Evan B. and {Dotter}, Aaron and {Conroy}, Charlie and {Cunningham}, Tim and {Park}, Minjung and {Tremblay}, Pier-Emmanuel},
        title = "{MESA Isochrones and Stellar Tracks (MIST). III. The White Dwarf Cooling Sequence}",
      journal = {\apjs},
         year = 2026,
        month = mar,
       volume = {283},
       number = {1},
          eid = {41},
        pages = {41},
          doi = {10.3847/1538-4365/ae401e},
archivePrefix = {arXiv},
       eprint = {2509.21717},
 primaryClass = {astro-ph.SR},
       adsurl = {https://ui.adsabs.harvard.edu/abs/2026ApJS..283...41B}
}

@ARTICLE{bedard2020,
       author = {{B{\'e}dard}, A. and {Bergeron}, P. and {Brassard}, P. and {Fontaine}, G.},
        title = "{On the Spectral Evolution of Hot White Dwarf Stars. I. A Detailed Model Atmosphere Analysis of Hot White Dwarfs from SDSS DR12}",
      journal = {\apj},
         year = 2020,
        month = oct,
       volume = {901},
       number = {2},
          eid = {93},
        pages = {93},
          doi = {10.3847/1538-4357/abafbe},
archivePrefix = {arXiv},
       eprint = {2008.07469},
 primaryClass = {astro-ph.SR},
       adsurl = {https://ui.adsabs.harvard.edu/abs/2020ApJ...901...93B}
}

@ARTICLE{althaus2013,
       author = {{Althaus}, Leandro G. and {Miller Bertolami}, Marcelo M. and {C{\'o}rsico}, Alejandro H.},
        title = "{New evolutionary sequences for extremely low-mass white dwarfs. Homogeneous mass and age determinations and asteroseismic prospects}",
      journal = {\aap},
         year = 2013,
        month = sep,
       volume = {557},
          eid = {A19},
        pages = {A19},
          doi = {10.1051/0004-6361/201321868},
archivePrefix = {arXiv},
       eprint = {1307.1882},
 primaryClass = {astro-ph.SR},
       adsurl = {https://ui.adsabs.harvard.edu/abs/2013A&A...557A..19A}
}

@ARTICLE{camisassa2016,
       author = {{Camisassa}, Mar{\'\i}a E. and {Althaus}, Leandro G. and {C{\'o}rsico}, Alejandro H. and {Vinyoles}, N{\'u}ria and {Serenelli}, Aldo M. and {Isern}, Jordi and {Miller Bertolami}, Marcelo M. and {Garc{\'\i}a{\textendash}Berro}, Enrique},
        title = "{The Effect of $^{22}$NE Diffusion in the Evolution and Pulsational Properties of White Dwarfs with Solar Metallicity Progenitors}",
      journal = {\apj},
         year = 2016,
        month = jun,
       volume = {823},
       number = {2},
          eid = {158},
        pages = {158},
          doi = {10.3847/0004-637X/823/2/158},
archivePrefix = {arXiv},
       eprint = {1604.01744},
 primaryClass = {astro-ph.SR},
       adsurl = {https://ui.adsabs.harvard.edu/abs/2016ApJ...823..158C}
}

@ARTICLE{camisassa2017,
       author = {{Camisassa}, Mar{\'\i}a E. and {Althaus}, Leandro G. and {Rohrmann}, Ren{\'e} D. and {Garc{\'\i}a-Berro}, Enrique and {Torres}, Santiago and {C{\'o}rsico}, Alejandro H. and {Wachlin}, Felipe C.},
        title = "{Updated Evolutionary Sequences for Hydrogen-deficient White Dwarfs}",
      journal = {\apj},
         year = 2017,
        month = apr,
       volume = {839},
       number = {1},
          eid = {11},
        pages = {11},
          doi = {10.3847/1538-4357/aa6797},
archivePrefix = {arXiv},
       eprint = {1703.05340},
 primaryClass = {astro-ph.SR},
       adsurl = {https://ui.adsabs.harvard.edu/abs/2017ApJ...839...11C}
}

@ARTICLE{camisassa2019,
       author = {{Camisassa}, Mar{\'\i}a E. and {Althaus}, Leandro G. and {C{\'o}rsico}, Alejandro H. and {De Ger{\'o}nimo}, Francisco C. and {Miller Bertolami}, Marcelo M. and {Novarino}, Mar{\'\i}a L. and {Rohrmann}, Ren{\'e} D. and {Wachlin}, Felipe C. and {Garc{\'\i}a-Berro}, Enrique},
        title = "{The evolution of ultra-massive white dwarfs}",
      journal = {\aap},
         year = 2019,
        month = may,
       volume = {625},
          eid = {A87},
        pages = {A87},
          doi = {10.1051/0004-6361/201833822},
archivePrefix = {arXiv},
       eprint = {1807.03894},
 primaryClass = {astro-ph.SR},
       adsurl = {https://ui.adsabs.harvard.edu/abs/2019A&A...625A..87C}
}

@ARTICLE{fontaine2001,
       author = {{Fontaine}, G. and {Brassard}, P. and {Bergeron}, P.},
        title = "{The Potential of White Dwarf Cosmochronology}",
      journal = {\pasp},
         year = 2001,
        month = apr,
       volume = {113},
       number = {782},
        pages = {409-435},
          doi = {10.1086/319535},
       adsurl = {https://ui.adsabs.harvard.edu/abs/2001PASP..113..409F}
}

\end{document}